\documentclass[onecolumn,superscriptaddress,showpacs,nofootinbib]{revtex4-2}

\usepackage[dvips]{graphicx} % for figures
\usepackage{amsfonts,amscd,amsmath,amsthm}
\usepackage{enumerate}
\usepackage{epsfig}
\usepackage{subfigure}
\usepackage{diagbox}
\usepackage{xcolor} 
\usepackage[colorlinks = true]{hyperref}
\usepackage{physics}
\usepackage{epstopdf}
\usepackage{framed}
\usepackage{multirow}
\usepackage{longtable} 
\usepackage{comment}
\usepackage[ruled,vlined]{algorithm2e}
\usepackage[most]{tcolorbox}
\usepackage{float} 
\usepackage{tabularx}
\usepackage{siunitx}
\usepackage{mathtools}
\usepackage{upgreek}
\usepackage{lipsum}
\usepackage{lineno}
\newtcolorbox[auto counter]{mybox}[2][]{
	enhanced,
	breakable,
	colback=blue!5!white,
	colframe=blue!75!black,
	fonttitle=\bfseries,
	title=Box \thetcbcounter: #2,#1
}

\begin{document}
%\linenumbers	

\title{Realization of an untrusted intermediate relay architecture using a quantum dot single-photon source}

\author{Mi Zou}
\thanks{Equal contribution.}
\affiliation{Hefei National Laboratory, University of Science and Technology of China, Hefei 230088, China}

\author{Yu-Ming He}
\thanks{Equal contribution.}
\affiliation{Hefei National Laboratory, University of Science and Technology of China, Hefei 230088, China}
\affiliation{Hefei National Research Center for Physical Sciences at the Microscale and School of Physical Sciences, University of Science and Technology of China, Hefei 230026, China}
\affiliation{New Cornerstone Science Laboratory, CAS Center for Excellence in Quantum Information and Quantum Physics, University of Science and Technology of China, Hefei 230026, China}

\author{Yizhi Huang}
\thanks{Equal contribution.}
\affiliation{Center for Quantum Information, Institute for Interdisciplinary Information Sciences, Tsinghua University, Beijing 100084, China}

\author{Jun-Yi Zhao}
\thanks{Equal contribution.}
\affiliation{Hefei National Laboratory, University of Science and Technology of China, Hefei 230088, China}
\affiliation{Hefei National Research Center for Physical Sciences at the Microscale and School of Physical Sciences, University of Science and Technology of China, Hefei 230026, China}
\affiliation{New Cornerstone Science Laboratory, CAS Center for Excellence in Quantum Information and Quantum Physics, University of Science and Technology of China, Hefei 230026, China}

\author{Bin-Chen Li}
\affiliation{Hefei National Laboratory, University of Science and Technology of China, Hefei 230088, China}
\affiliation{Hefei National Research Center for Physical Sciences at the Microscale and School of Physical Sciences, University of Science and Technology of China, Hefei 230026, China}
\affiliation{New Cornerstone Science Laboratory, CAS Center for Excellence in Quantum Information and Quantum Physics, University of Science and Technology of China, Hefei 230026, China}

\author{Yong-Peng Guo}
\affiliation{Hefei National Laboratory, University of Science and Technology of China, Hefei 230088, China}
\affiliation{Hefei National Research Center for Physical Sciences at the Microscale and School of Physical Sciences, University of Science and Technology of China, Hefei 230026, China}
\affiliation{New Cornerstone Science Laboratory, CAS Center for Excellence in Quantum Information and Quantum Physics, University of Science and Technology of China, Hefei 230026, China}

\author{Xing Ding}
\affiliation{Hefei National Laboratory, University of Science and Technology of China, Hefei 230088, China}

\author{Mo-Chi Xu}
\affiliation{Hefei National Laboratory, University of Science and Technology of China, Hefei 230088, China}
\affiliation{Hefei National Research Center for Physical Sciences at the Microscale and School of Physical Sciences, University of Science and Technology of China, Hefei 230026, China}
\affiliation{New Cornerstone Science Laboratory, CAS Center for Excellence in Quantum Information and Quantum Physics, University of Science and Technology of China, Hefei 230026, China}

\author{Run-Ze Liu}
\affiliation{Hefei National Laboratory, University of Science and Technology of China, Hefei 230088, China}
\affiliation{Hefei National Research Center for Physical Sciences at the Microscale and School of Physical Sciences, University of Science and Technology of China, Hefei 230026, China}
\affiliation{New Cornerstone Science Laboratory, CAS Center for Excellence in Quantum Information and Quantum Physics, University of Science and Technology of China, Hefei 230026, China}

\author{Geng-Yan Zou}
\affiliation{Hefei National Laboratory, University of Science and Technology of China, Hefei 230088, China}
\affiliation{Hefei National Research Center for Physical Sciences at the Microscale and School of Physical Sciences, University of Science and Technology of China, Hefei 230026, China}
\affiliation{New Cornerstone Science Laboratory, CAS Center for Excellence in Quantum Information and Quantum Physics, University of Science and Technology of China, Hefei 230026, China}

\author{Zhen Ning}
\affiliation{Hefei National Laboratory, University of Science and Technology of China, Hefei 230088, China}
\affiliation{Hefei National Research Center for Physical Sciences at the Microscale and School of Physical Sciences, University of Science and Technology of China, Hefei 230026, China}
\affiliation{New Cornerstone Science Laboratory, CAS Center for Excellence in Quantum Information and Quantum Physics, University of Science and Technology of China, Hefei 230026, China}

\author{Xiang You}
\affiliation{Hefei National Laboratory, University of Science and Technology of China, Hefei 230088, China}

\author{Hui Wang}
\affiliation{Hefei National Laboratory, University of Science and Technology of China, Hefei 230088, China}
\affiliation{Hefei National Research Center for Physical Sciences at the Microscale and School of Physical Sciences, University of Science and Technology of China, Hefei 230026, China}
\affiliation{New Cornerstone Science Laboratory, CAS Center for Excellence in Quantum Information and Quantum Physics, University of Science and Technology of China, Hefei 230026, China}

\author{Wen-Xin Pan}
\affiliation{Hefei National Laboratory, University of Science and Technology of China, Hefei 230088, China}
\affiliation{Hefei National Research Center for Physical Sciences at the Microscale and School of Physical Sciences, University of Science and Technology of China, Hefei 230026, China}
\affiliation{New Cornerstone Science Laboratory, CAS Center for Excellence in Quantum Information and Quantum Physics, University of Science and Technology of China, Hefei 230026, China}

\author{Hao-Tao Zhu}
\affiliation{Hefei National Laboratory, University of Science and Technology of China, Hefei 230088, China}
\affiliation{Hefei National Research Center for Physical Sciences at the Microscale and School of Physical Sciences, University of Science and Technology of China, Hefei 230026, China}
\affiliation{New Cornerstone Science Laboratory, CAS Center for Excellence in Quantum Information and Quantum Physics, University of Science and Technology of China, Hefei 230026, China}

\author{Ming-Yang Zheng}
\affiliation{Hefei National Laboratory, University of Science and Technology of China, Hefei 230088, China}
\affiliation{Jinan Institute of Quantum Technology and CAS Center for Excellence in Quantum Information and Quantum Physics, University of Science and Technology of China, Jinan 250101, China}

\author{Xiu-Ping Xie}
\affiliation{Hefei National Laboratory, University of Science and Technology of China, Hefei 230088, China}
\affiliation{Jinan Institute of Quantum Technology and CAS Center for Excellence in Quantum Information and Quantum Physics, University of Science and Technology of China, Jinan 250101, China}

\author{Dandan Qin}
\affiliation{Hefei National Laboratory, University of Science and Technology of China, Hefei 230088, China}
\affiliation{Hefei National Research Center for Physical Sciences at the Microscale and School of Physical Sciences, University of Science and Technology of China, Hefei 230026, China}
\affiliation{New Cornerstone Science Laboratory, CAS Center for Excellence in Quantum Information and Quantum Physics, University of Science and Technology of China, Hefei 230026, China}

\author{Xiao Jiang}
\affiliation{Hefei National Laboratory, University of Science and Technology of China, Hefei 230088, China}
\affiliation{Hefei National Research Center for Physical Sciences at the Microscale and School of Physical Sciences, University of Science and Technology of China, Hefei 230026, China}
\affiliation{New Cornerstone Science Laboratory, CAS Center for Excellence in Quantum Information and Quantum Physics, University of Science and Technology of China, Hefei 230026, China}

\author{Yong-Heng Huo}
\affiliation{Hefei National Laboratory, University of Science and Technology of China, Hefei 230088, China}
\affiliation{Hefei National Research Center for Physical Sciences at the Microscale and School of Physical Sciences, University of Science and Technology of China, Hefei 230026, China}
\affiliation{New Cornerstone Science Laboratory, CAS Center for Excellence in Quantum Information and Quantum Physics, University of Science and Technology of China, Hefei 230026, China}

\author{Qiang Zhang}
\affiliation{Hefei National Laboratory, University of Science and Technology of China, Hefei 230088, China}
\affiliation{Hefei National Research Center for Physical Sciences at the Microscale and School of Physical Sciences, University of Science and Technology of China, Hefei 230026, China}
\affiliation{Jinan Institute of Quantum Technology and CAS Center for Excellence in Quantum Information and Quantum Physics, University of Science and Technology of China, Jinan 250101, China}

\author{Chao-Yang Lu}
\email{cylu@ustc.edu.cn}
\affiliation{Hefei National Laboratory, University of Science and Technology of China, Hefei 230088, China}
\affiliation{Hefei National Research Center for Physical Sciences at the Microscale and School of Physical Sciences, University of Science and Technology of China, Hefei 230026, China}
\affiliation{New Cornerstone Science Laboratory, CAS Center for Excellence in Quantum Information and Quantum Physics, University of Science and Technology of China, Hefei 230026, China}

\author{Xiongfeng Ma}
\email{xma@tsinghua.edu.cn}
\affiliation{Center for Quantum Information, Institute for Interdisciplinary Information Sciences, Tsinghua University, Beijing 100084, China}
\affiliation{Hefei National Laboratory, University of Science and Technology of China, Hefei 230088, China}

\author{Teng-Yun Chen}
\email{tychen@ustc.edu.cn}
\affiliation{Hefei National Laboratory, University of Science and Technology of China, Hefei 230088, China}
\affiliation{Hefei National Research Center for Physical Sciences at the Microscale and School of Physical Sciences, University of Science and Technology of China, Hefei 230026, China}
\affiliation{New Cornerstone Science Laboratory, CAS Center for Excellence in Quantum Information and Quantum Physics, University of Science and Technology of China, Hefei 230026, China}

\author{Jian-Wei Pan}
\email{pan@ustc.edu.cn}
\affiliation{Hefei National Laboratory, University of Science and Technology of China, Hefei 230088, China}
\affiliation{Hefei National Research Center for Physical Sciences at the Microscale and School of Physical Sciences, University of Science and Technology of China, Hefei 230026, China}
\affiliation{New Cornerstone Science Laboratory, CAS Center for Excellence in Quantum Information and Quantum Physics, University of Science and Technology of China, Hefei 230026, China}

\begin{abstract}
To fully exploit the potential of quantum technologies, quantum networks are needed to link different systems, significantly enhancing applications in computing, cryptography, and metrology. Central to these networks are quantum relays that can facilitate long-distance entanglement distribution and quantum communication. In this work, we present a modular and scalable quantum relay architecture using a high-quality single-photon source. The proposed network incorporates three untrusted intermediate nodes and is capable of a repetition rate of 304.52 MHz. We use a measurement-device-independent protocol to demonstrate secure key establishment over fibers covering up to 300 kilometers. This study highlights the potential of single-photon sources in quantum relays to enhance information transmission, expand network coverage, and improve deployment flexibility, with promising applications in future quantum networks.
\end{abstract}

\maketitle 

\section{Introduction}
Quantum networks have emerged as a cornerstone of quantum information science, with the potential to interconnect diverse quantum systems and revolutionize areas such as quantum computing \cite{shor1994algorithms,grover1996a}, cryptography \cite{bennett1984quantum,ekert1991Quantum}, and metrology \cite{giovannetti2004quantum,giovannetti2011advances}. Recent progress includes the construction of large-scale quantum wide-area networks, such as the Beijing-Shanghai backbone \cite{Chen2021integrated}, and broader global initiatives to realize scalable quantum communication systems \cite{peev2009secoqc, sasaki2011field}.

Photonic quantum information transmission offers seamless integration with existing optical fiber networks \cite{simon2017towards}, enabling the connection of diverse quantum computing platforms and enhancing their computational capabilities \cite{van2016path,broadbent2009universal}. However, current quantum networks often rely on classical relay nodes \cite{Chen2021integrated}, which convert quantum information into classical data for processing. This approach inherently limits their ability to transmit quantum information or preserve entanglement. Moreover, classical relays require complete trust, as they have unrestricted access to the transmitted data. As quantum networks evolve to directly connect quantum devices, classical relays fall short in managing essential quantum resources, such as entanglement, underscoring the need for more advanced quantum relay technologies.

To address these limitations, quantum repeaters were proposed \cite{Briegel1998Repeaters,duan2001long}. Quantum repeaters establish entanglement across multiple nodes, forming the foundation for advanced quantum protocols and distributed quantum computing. However, practical implementation remains challenging due to their reliance on complex technologies like entanglement distillation, quantum memories, and non-demolition measurements, which are not yet viable in the near term.

As an alternative, quantum relays offer a simpler solution \cite{jacobs2002quantum,waks2002security,de2004long,collins2005quantum}. Quantum relays operate similarly to repeaters but avoid the need for intricate quantum operations or devices. Unlike classical relays, quantum relays function as untrusted nodes in cryptographic systems, leveraging quantum properties like superposition and entanglement to securely and efficiently transmit quantum information. By integrating quantum relay nodes into network infrastructures, quantum networks become more practical, flexible, and adaptable to diverse communication scenarios and application requirements.

There have been numerous efforts to implement quantum relays \cite{jing2019entanglement,Yu2020Entanglement,pompili2021realization,hermans2022qubit,liu2024creation}, but these implementations are often limited to a single quantum relay node positioned centrally. However, scaling quantum networks to include multiple untrusted relay nodes is essential for advancing their development. Achieving this scalability poses significant challenges, including synchronization and interference management. Existing structures frequently struggle to scale efficiently while maintaining security and reliability, underscoring the need for innovative relay designs that can support multiple untrusted nodes and enhance the robustness and scalability of quantum networks.

\begin{figure}[t!]
	\centering 
	\includegraphics[width=0.8\linewidth]{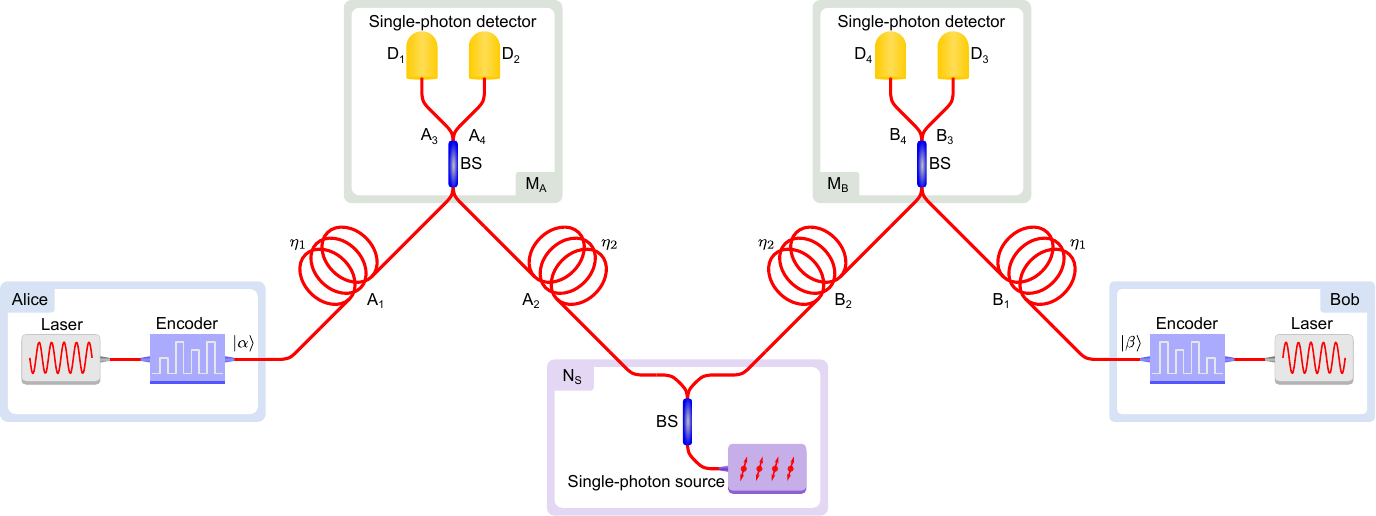}
	\caption{Five-node network structure. The structure comprises two end-user nodes (Alice and Bob), two measurement nodes (M\textsubscript{A} and M\textsubscript{B}), and a central quantum relay node housing the single-photon source (N\textsubscript{S}). The distances or losses between the nodes are not necessarily equal. The single photons from the source are transmitted via paths $A_2$ and $B_2$ to the measurement nodes after passing through a 50:50 beam splitter (BS). Here they interfere with the coherent light pulses from the user nodes. The interference results are then detected by single-photon detectors (paths $A_3$/$A_4$/$B_3$/$B_4$). Channel transmittances are denoted by $\eta_1$ (user measurement) and $\eta_2$ (N\textsubscript{S} measurement). Single-photon pulses arriving at the two measurement nodes are entangled along their paths. 
The measurement process resembles quantum teleportation or entanglement swapping, enabling quantum information transmission between the two user nodes. In our setup, the single-photon source and both measurement nodes are integral to establish quantum correlations between the users. Consequently, all three intermediate nodes are considered quantum relays within our structure. Moreover, similar to the measurement-device-independent protocol, this entanglement swapping process allows the three central nodes to be completely untrusted in this network structure.}
\label{fig:protocol}
\end{figure}
%%%%five-node structure%%%%
In this study, we present a five-node quantum network structure, as illustrated in Fig.~\ref{fig:protocol}. The two end-users, Alice and Bob, employ identical encoders to encode quantum information onto light pulses. While phase encoding is utilized in our experiment, other optical modes, such as intensity or polarization, could also be adopted. A centrally positioned single-photon source acts as a quantum relay, emitting single-photon pulses that are split and directed toward Alice and Bob. At the measurement nodes, M\textsubscript{A} and M\textsubscript{B}, these single photons interfere with the light pulses sent by Alice and Bob, and the interference results are detected by single-photon detectors.

The five-node relay design introduces three intermediate nodes, the single-photon source and the two measurement nodes, functioning as untrusted quantum nodes. This structure supports the creation of complex network topologies, such as three-layer star configurations, while enhancing the signal-to-noise ratio of individual links. By improving robustness and extending the maximum communication distance, this design addresses key challenges in quantum network scalability and performance. Moreover, the untrusted nature of the intermediate nodes removes the need for stringent security and placement constraints, enabling cost-effective deployment in dynamic environments. This modular approach facilitates implementation in a wide range of communication scenarios and underscores the practicality of the proposed structure.

Quantum information transfer within this five-node setup requires coincident detection events at the two measurement nodes. Users must post-select results to identify successful relays. Under ideal conditions, ignoring losses in the channel and phase perturbations in the fiber, the post-selected states corresponding to the four detector paths after interference are expressed as (unnormalized):
\begin{equation}\label{eq:phasetran}
	\sqrt{\eta_1\eta_2}(\alpha+\beta)(a_{3}^\dagger b_{3}^\dagger + a_{4}^\dagger b_{4}^\dagger)\ket{0} + \sqrt{\eta_1\eta_2}(\alpha -\beta)(a_{3}^\dagger b_{4}^\dagger + a_{4}^\dagger b_{3}^\dagger)\ket{0},
\end{equation}
where $\eta_1$ and $\eta_2$ denote the transmittances from the users and the single-photon source to the measurement nodes, respectively. $\alpha$ and $\beta$ correspond to the coherent states $\ket{\alpha}$ and $\ket{\beta}$ sent by Alice and Bob, and $a_3^\dagger$, $a_4^\dagger$, $b_3^\dagger$, $b_4^\dagger$ are the creation operators for the four detector paths.

Consider the case where $|\alpha| = |\beta|$. When the phases of $\ket{\alpha}$ and $\ket{\beta}$ are identical, either $D_1$ and $D_3$, or $D_2$ and $D_4$, will click simultaneously. Conversely, when the phase difference is $\pi$, the simultaneous clicks will occur between $D_1$ and $D_4$, or $D_2$ and $D_3$. This allows the users to infer the phase relationship of the other party based on the detection results, thereby achieving quantum information transfer using the single-photon source. In our setup, both the single-photon source and the interference-based measurement nodes are indispensable to this process, and we therefore regard all three intermediate nodes as quantum relays. A more detailed analysis and derivation are provided in the Appendix \ref{sc:protocol}.

To further demonstrate the practicality of the proposed five-node structure, we implement a variant of the twin-field quantum key distribution (QKD) protocol \cite{lucamarini2018overcoming} --- phase-matching scheme \cite{Ma2018phase} with the decoy-state method \cite{hwang2003decoy,Lo2005Decoy,wang2005decoy} --- to successfully transmit secure information between two users. The successful realization of QKD also confirms that the setup establishes a quantum channel capable of distributing distillable entanglement \cite{lo1999Unconditional}. As a result, this structure can serve as a relay to distribute entanglement for future quantum networks.

%%%%%%%%SPS introduction%%%%%%%%

\section{Single photon source and its interference}
\begin{figure}[t!]
	\centering 
	\includegraphics[width=0.8\linewidth]{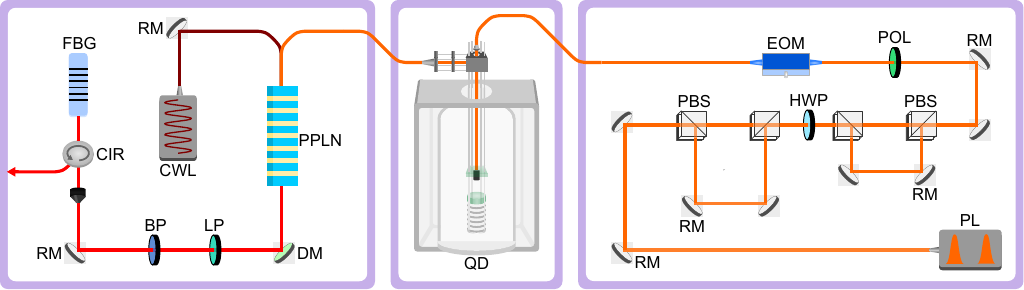}
\caption{Single-photon source setup. An InAs/GaAs quantum dot (QD) coupled to a tunable plane-concave Fabry-Perot cavity emits single photons at a wavelength of $\SI{884.5}{\nm}$ when resonantly excited by a pulsed laser (PL). The PL generates light pulses with a repetition rate of $\SI{76.13}{\MHz}$, which can be quadrupled to $\SI{304.52}{\MHz}$ through beam splitting and combining twice with half-cycle delay differences. Some of these pulses can be selectively eliminated using an electro-optic modulator (EOM). The wavelength of the single photon is converted to $\SI{1550.2}{\nm}$ using a periodically-poled lithium niobate (PPLN) waveguide pumped by a $\SI{2059.7}{\nm}$ continuous-wave laser (CWL). The single photons are filtered by a dichromatic mirror (DM), a long pass (LP), a bandpass (BP), and a fiber Bragg grating (FBG) with a bandwidth of $\SI{5}{\GHz}$ to improve purity. CIR, circulator; RM, reflective mirror; PBS, polarization BS; HWP, half-wave plate; POL, polarizer.}
\label{fig:sps}
\end{figure}
Central to our quantum relay structure is the single-photon source, a critical component for ensuring high-quality interference and effective quantum information relay. Extensive research has explored various single-photon source systems, including color centers in diamonds \cite{kurtsiefer2000stable, aharonovich2016solid}, trapped ions \cite{keller2004continuous}, quantum dots \cite{michler2000quantum, ding2016demand, ding2023highefficiency}, and atom ensembles \cite{chou2004single, ripka2018room}. Among these, semiconductor quantum dot single-photon sources stand out due to their high single-photon efficiency, indistinguishability, and high repetition rate. They can efficiently generate entanglement along photon paths, enabling quantum information relay through Bell-state measurements in multi-node structures. Their compatibility with room temperature operation and minimal environmental interactions make them highly practical for real-world applications. Additionally, quantum-dot-based sources provide a scalable, solid-state platform with ultrabrightness and seamless integration with matter qubits. Given these advantages, we utilize a high-purity quantum dot single-photon source \cite{ding2023highefficiency} in our relay setup.

The setup of our single-photon source is illustrated in Fig.~\ref{fig:sps}. A pulse laser with a wavelength of $\SI{884.5}{\nm}$ generates narrow light pulses at a frequency of $\SI{76.13}{\MHz}$, which can be quadrupled to $\SI{304.52}{\MHz}$ through two asymmetrical Mach-Zehnder interferometer. A single InAs/GaAs quantum dot, coupled to a tunable plane-concave Fabry-Perot cavity, is resonantly excited by these light pulses, leading to the emission of single photons. To ensure selective photon emission, we have incorporated an electro-optic modulator. This allows us to eliminate specific excitation pulses, preventing the quantum dot from emitting single photons at desired pulses. To enable interference with the coherent light pulses from the user nodes and ensure compatibility with existing optical fiber systems, the single-photon wavelength is converted to the telecommunication wavelength of $\SI{1550.2}{\nm}$ using a periodically-poled lithium niobate waveguide, pumped by a $\SI{2059.7}{\nm}$ continuous-wave laser. The purity of single photons is enhanced by a few filters including a dichromatic mirror, a long pass, a bandpass, and a fiber Bragg grating with a bandwidth of $\SI{5}{\GHz}$. 

\begin{figure}[htbp!]
	\centering 
	\includegraphics[width=\linewidth]{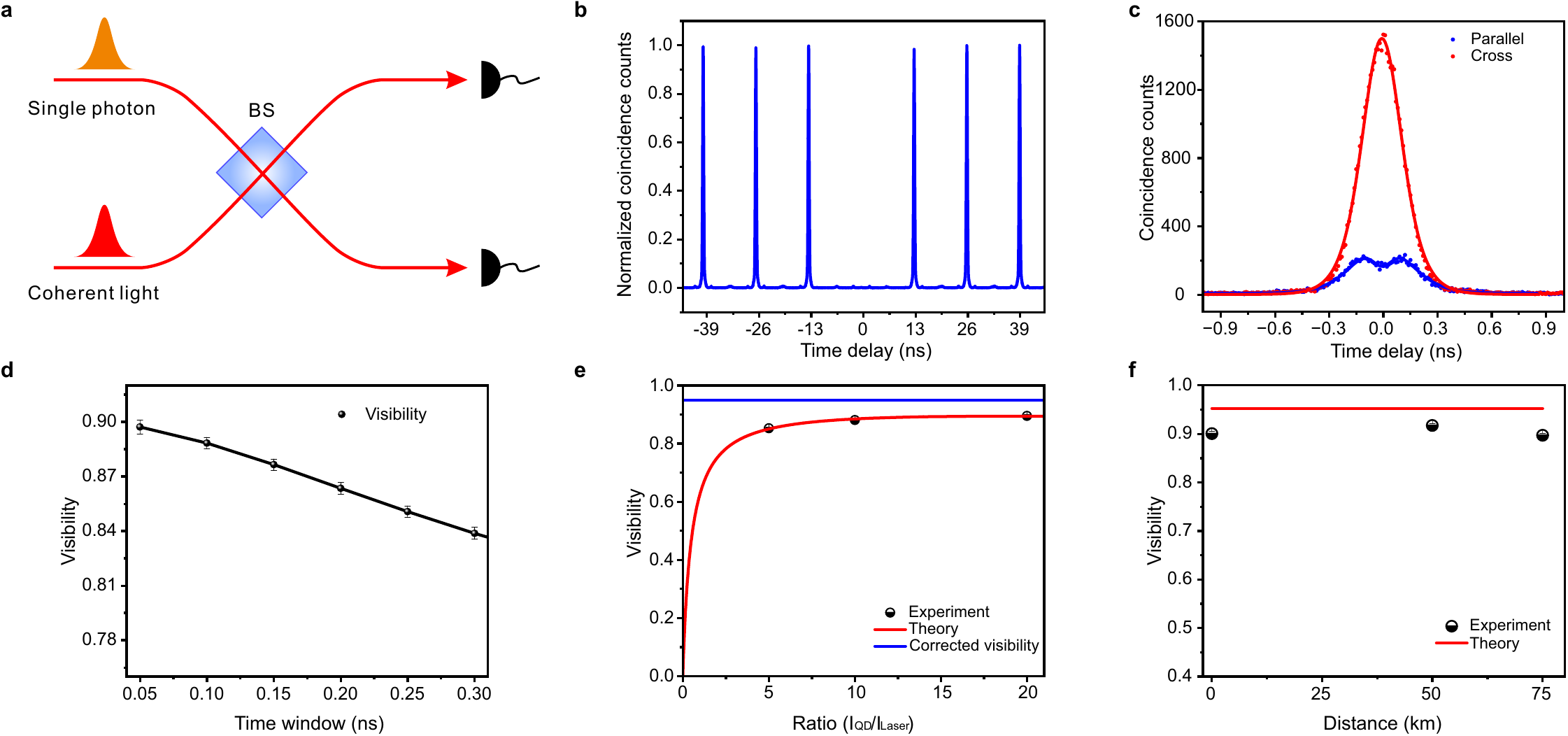}
	\caption{Interference between single photons and coherent light pulses. $\textbf{a,}$ Schematic of the interference between coherent light pulses and single photons. $\textbf{b,}$ Normalized coincidence counts when only single photons are emitted. The $g^{(2)}(0)$ of the single-photon source is fitted based on this result. $\textbf{c,}$ Coincidence counts $c_1$ and $c_2$ recorded at the same and different wavelengths, represented by blue and red dots, respectively. The blue and red lines indicate the corresponding fitting results. The intensity ratio between coherent light pulses and single photons is 1:10. $\textbf{d,}$ Interference visibility as a function of the detection time window. The visibility is calculated by $1-c_1/c_2$. $\textbf{e,}$ Interference visibility versus intensity ratio. The red line and black dots represent theoretical and experimental results, respectively. The blue line indicates the corrected visibility. The detection time window is $\SI{50}{\ps}$. $\textbf{f,}$ Interference visibility as a function of transmission distance. The red line and black dots represent theoretical and experimental results, respectively. The intensity ratio between coherent light pulses and single photons is 1:10. The detection time window is $\SI{50}{\ps}$. Data points in $\textbf{d}$–$\textbf{f}$ are extracted from model fits to experimental data (e.g., $\textbf{c}$). Error bars represent fitting uncertainties (standard errors).}\label{fig:visibility}
\end{figure}

In our setup, the relay of quantum information is primarily achieved through interference. Hence, ensuring the quality of interference between the single photon and the coherent light --- essentially, the indistinguishability of the photons --- is crucial for the success of our relay. To do this, we first measure the interference visibility between single photons and coherent light pulses at the two measurement nodes. The schematic of the test is shown in Fig.~\ref{fig:visibility}$\textbf{a}$. Pulses emitted by the single-photon source and the laser source interfere at a balanced beamsplitter and are then detected by two single-photon detectors. The visibility $V$ is defined by $V=1-c_1/c_2$, where $c_1$ and $c_2$ represent coincidence counts when the input lights have the same and different wavelengths, respectively. The closer the visibility is to 1, the higher the quality of the interference. The maximum interference visibility between the coherent pulse and the single photon depends on several factors, including the single-photon intrinsic indistinguishability, single-photon purity, spectral overlap between the single-photon and coherent light, and the intensity ratio between the single-photon and coherent light. The single-photon intrinsic indistinguishability is determined by the single-photon source, and the indistinguishability of the source is about 98.6\%~\cite{ding2023highefficiency}.

To assess the impact of the remaining factors and maximize interference visibility, we conducted a series of tests. We first record the coincidence counts when the coherent light pulses are blocked, allowing only the single photons to be emitted, as shown in Fig.~\ref{fig:visibility}$\textbf{b}$. By fitting this result, we determine the second-order correlation function of the emitted light from our single-photon source to be
\begin{equation}
	g^{(2)}(0)=\dfrac{\langle n(n-1)\rangle}{\langle n^2 \rangle}=0.0015,
\end{equation}
where $n$ represents the sum of the probabilities of the two detectors clicking in the interference setup. This result indicates that our light source has extremely high purity, as a perfect single-photon source would have $g^{(2)}(0)=0$.

We then make the coherent pulse and the single photon arrive at the beam splitter simultaneously for interference. With the intensity ratio between the coherent light pulses and single photons set to 1:10, we first record the coincidence counts $c_1$ and $c_2$, represented by blue and red dots in Fig.~\ref{fig:visibility}$\textbf{c}$, respectively. This result confirms the effectiveness of the interference between the single photon and the coherent light in our system. 

Building on this, we aim to achieve higher interference visibility by optimizing key experimental parameters. One critical factor we considered is the detection time window, as $c_1$ and $c_2$ are highly sensitive to its configuration. The underlying issue stems from the spectral overlap between the single photon and the coherent light, which directly affects the interference. To address this, we simplify the spectral overlap into two factors: the central wavelength and the pulse shape. The laser wavelength is carefully tuned to match that of the single photon. However, due to fundamental differences in the mechanisms of single-photon generation and coherent pulse production, their pulse shapes are inherently distinct. To reconcile this discrepancy, we optimize the detection time window. To evaluate its effect, we measured interference visibility under different detection time windows while maintaining a fixed intensity ratio. The results, shown in Fig.~\ref{fig:visibility}$\textbf{d}$, were fitted to highlight the trend. The data confirm that narrowing the detection time window enhances the spectral overlap between the single photon and the coherent light, leading to a significant improvement in interference visibility.

Another factor influencing the interference visibility is the intensity ratio between the coherent light and the single photon. Typically, optimal visibility in interference experiments requires matching intensities of identical light sources. However, in our experimental setup involving a single photon light source and a laser light source, this principle does not directly apply. We test the interference visibility across varying intensity ratios, with the experimental results depicted by black dots and the theoretical predictions represented by a red line in Fig.~\ref{fig:visibility}$\textbf{e}$. Here, the detection time window is fixed at $\SI{50}{\ps}$. Additionally, we calculate a theoretically corrected visibility, indicated by the solid blue line, which accounts for conditions excluding multi-photon events and imperfections such as detector dark counts, reflecting the ideal indistinguishability between single photons and coherent light photons. The derivation of the theoretical results here can be found in the Appendix \ref{sc:protocol}. These results demonstrate that, to achieve optimal visibility, the intensity of the coherent state must be significantly lower than that of the single photon. As the intensity ratio between the coherent pulse and the single photon decreases, the interference visibility correspondingly improves.

Finally, we investigate the effect of communication fiber length on interference visibility, with results presented in Fig.~\ref{fig:visibility}$\textbf{f}$. The black dots represent experimental data, while the red line corresponds to theoretical calculations that account for factors such as bandwidth mismatch between the coherent light pulse and the single photon, as well as the imperfect $g^{(2)}(0)$ of the single-photon source. The theoretical corrected two-photon interference visibility is 0.95. The test results confirm that as the distance increases, interference visibility remains consistently high, ensuring the scalability of our relay structure across extended distances.

These interference tests demonstrate that in our system, by carefully optimizing the intensity ratio and detection time window, the interference visibility between the single-photon source and the laser source can reach values exceeding 0.85, and even approach 0.9. This high-quality interference meets the stringent requirements necessary for reliable quantum information relay in our five-node structure. While the test results show that reducing the coherent pulse intensity and narrowing the detection time window improves interference visibility, it simultaneously decreases the number of detection events, which reduces the overall system efficiency. Therefore, after balancing the trade-off between visibility and efficiency, we chose an intensity ratio of approximately 1:10 and a detection time window of $\SI{200}{\ps}$ for the subsequent experiments.

\section{Quantum communication scheme with three untrusted nodes}
After ensuring high-quality interference, we then proceeded to implement a QKD protocol in our five-node structure to effectively transmit information and further verify the feasibility of this relay structure. Considering that interference directly reflects the phase information of quantum states, we employed a modified phase-matching QKD scheme and we use the key rate generated serves as a measure of our system's ability to relay quantum information and distribute entanglement.

Here, we provide an overview of the modified phase-matching scheme in Box \ref{box:PMQKD}. The experimental setting at the user nodes and measurement nodes to realize the scheme are presented in Methods.
%Details of encoding setting and phase estimation technique are presented in the Methods.
\begin{mybox}[label={box:PMQKD}]{Phase-matching QKD scheme with three untrusted nodes}
	\begin{enumerate}[(1)]
		\item 
		State Preparation: Alice prepares coherent state $\ket{\alpha}=\ket{\sqrt{\mu}e^{ \mathrm{i} (\pi\kappa_a+\phi_a)}}_{a_1}$ on optical mode $a_1$, where $\kappa_a$ is Alice's raw key bit and phase $\phi_a$ is uniformly chosen from $\{0, 2\pi/D, 4\pi/D, \dots, 2(D-1)\pi/D\}$ with $D=16$. 
		Similarly, Bob randomly chooses $\kappa_b$ and $\phi^i$, and prepares $\ket{\beta_i}=\ket{\sqrt{\mu}e^{ \mathrm{i} (\pi\kappa_b+\phi_b)}}_{b_1}$ on mode $b_1$.
		At the relay node, a single-photon pulse is generated by the single-photon source and passes through a 50:50 beam splitter. The outgoing paths are denoted as optical modes $a_2$ and $b_2$.
		
		\item
		Measurement: The user nodes and the relay node send their optical modes to untrusted measurement nodes. The measurement nodes are supposed to perform single-photon interference measurement and record the clicks of detectors $L$ and/or $R$.

		\item 
		Announcement: For all rounds with successful detection, the two measurement nodes announce the $L$ and $R$ detection results.
		If both detection nodes announce successful detection within the same round, then Alice and Bob consider that round to be a valid round. Then, Alice and Bob announce the random phases, $\phi_a$ and $\phi_b$, of all valid rounds.
		
		\item 
		Key mapping: For those valid rounds, Alice and Bob compare their encoded random phases $\phi_a$ and $\phi_b$. If $\abs{\phi_a-\phi_b}=0$ or $\pi$, Alice and Bob keep the corresponding $\kappa_a$ and $\kappa_b$ as their raw key bits in this round, respectively. Otherwise, the discard their raw key bits. Additionally, Bob flips his key bit $\kappa_b^i$ if the announcement of two measurement nodes are different and he also flip his key bit if $\abs{\phi^i_a-\phi^i_b}=\pi$.
		
		\item 
		Post-processing: After getting the raw key bits, Alice and Bob do the parameter estimation, information reconciliation, and privacy amplification to get the final key bits.
		
	\end{enumerate}
\end{mybox}
Here we define key rate $r$ as the proportion of final key bits that can be extracted from each raw key bit. Then, the final key rate is \cite{Ma2018phase,Zeng2019Symmetryprotected}
\begin{equation}\label{eq:keyR}
	r = 1 - h\left(e_p\right) -f h\left(E_z \right),
\end{equation}
where $h(x)=-x \log_2 x-(1-x)\log_2(1-x)$ is binary entropy function, $E_Z$ is the bit error rate, $e_p$ is the phase error rate, and $f$ is the error correction efficiency. From a security perspective, since the phase-matching protocol is measurement-device-independent, considering the single-photon source and the two measurement nodes as a single measurement unit, the security of our system aligns with the original phase-matching protocol. For a detailed protocol description with decoy state method and performance analysis, please refer to the Appendix \ref{sc:protocol}. 

The experimental parameters and results of the phase-matching scheme are listed in Table \ref{tab:expres}. The total distances of $\SI{100}{\km}$, $\SI{200}{\km}$, and $\SI{300}{\km}$ are composed of four fiber links of $\SI{25}{\km}$, $\SI{50}{\km}$, and $\SI{75}{\km}$, respectively. In this symmetrical setup, the intensity of the signal state is much weaker than that optimized by conventional phase-matching scheme to ensure sufficiently high visibility. We set the signal state to an intensity of about 0.002 to balance the error rate and the effective click rate. It is worth mentioning that in the asymmetric case, where the distance between Alice or Bob and the measurement node is greater than the distance between the quantum relay node and the measurement node, the transmitted light intensity of Alice and Bob can be increased to maintain the same intensity ratio between the coherent states and the single photons, thus achieving the same effective click rate.

\begin{figure}[ht!]
	\includegraphics[width=8cm]{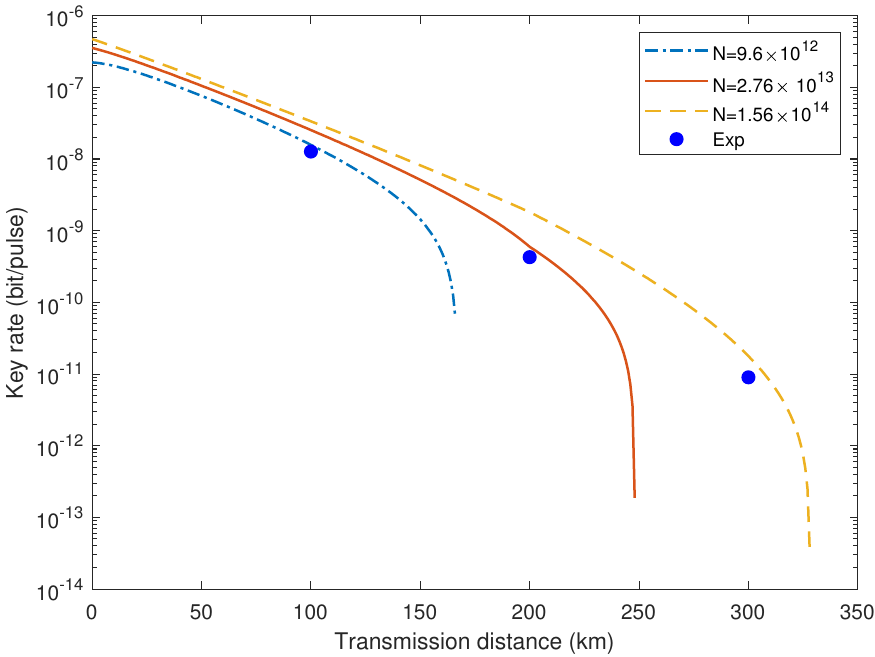}
	\caption{Key-rate performance. The points in the figure represent the experimental results, whereas the different curves correspond to the simulated key rates under different total rounds. Note that these curves were not optimized for different distances, but instead, used the intensity and transmission ratios consistent with actual experimental settings. These three curves use $\mu = 0.002$, $\nu = 0.001$, and the transmission ratios were obtained by fitting the parameters from Table \ref{tab:expres}. The curves exhibit slight discontinuities around the 200-km mark because different fitting results are used before and after 200 km to ensure that the simulation parameters align with the actual experimental parameters.}
	\label{fig:keyrate}
\end{figure}

The key rates for different transmission distances are presented in Fig.~\ref{fig:keyrate}, along with the simulation curve. Our system achieves key rates of $1.27\times10^{-8}$ bit/pulse, $4.27\times10^{-10}$ bit/pulse, and $8.99\times10^{-12}$ bit/pulse at distances of $\SI{100}{\km}$, $\SI{200}{\km}$, and $\SI{300}{\km}$, respectively. At these distances, our system exhibits quantum bit error rates of approximately 9.70\%, 9.57\%, and 10.44\%, respectively. The sources of these error rates can be divided into two aspects. Firstly, achieving high photon indistinguishability between coherent states and single photons remains a considerable experimental challenge. Due to differences in central wavelength and pulse shape, the interference visibility is reduced, which in turn leads to an increase in bit error rate—approximately 4–5\% in our case. This suggests potential for improving the interference visibility, thereby reducing the bit error rate and enhancing overall system performance. The relationship between interference visibility and the error rate is provided in the Appendix \ref{sc:tpi}. Secondly, the phase modulation procedure also influences the error rates. Phase modulation can be divided into active phase modulation and passive phase modulation. Active phase modulation is carried out using a phase modulator, and the inaccuracies in the modulation signal give rise to roughly 1\% bit error rate. Passive phase modulation involves phase shifts caused by the frequency difference between Alice's and Bob's lasers and the phase drift of the fiber links due to environmental influences. We utilize the optical phase-locked loop technique to phase-lock Alice's and Bob's lasers and compensates for drifting phases with phase reference pulses. As the transmission distance extends, phase noise stemming from locking imperfections and compensation errors due to delayed tracking accumulates gradually, causing the error rate to increase further.

\begin{table}[h]
	\centering
	\caption{\label{tab:expres} Experimental parameters for different distances $L$. The total number of optical pulses for QKD sent by users is $N$. The average intensities of signal and decoy states are denoted as $\mu$ and $\nu$, respectively, and the corresponding probabilities of sending these states are denoted as $p_\mu$ and $p_\nu$, respectively. The detector dark count is about $\SI{60}{\Hz}$. The average efficiency of the four single-photon detectors is about $82\%$. The total detection efficiency $\eta_d$ is about $50\%$, due to the $\SI{1.10}{\dB}$ insertion loss and the non-overlapping of the signal pulse with the detection time window. The latter is affected by time synchronization, and the greater the time jitter, the worse the overlap between the signal pulse and the detection time window. So, there is a reduction of the total detection efficiency at $\SI{300}{\km}$.}
	\begin{tabular}{l|lll}
		\hline 
		\hline 
		$L$   		& $\SI{100}{\km}$  	& $\SI{200}{\km}$ 	& $\SI{300}{\km}$ \\  
		\hline 
		$N$ 		&$9.6\times10^{12}$ &$2.76\times10^{13}$&$1.56\times10^{14}$  \\ 
		$\eta_d$ 	&0.52 				&0.52 				&0.48\\
		$\eta_1$ 	&0.325 				&0.130 				&0.047\\
		$\eta_2$ 	&0.332 				&0.127 				&0.051\\
		$\mu$ 		&0.00199 			&0.00199 			&0.00213\\
		$\nu$ 		&0.00080 			&0.00098 			&0.00103\\
		$p_\mu$		&0.751	 			&0.618 				&0.269\\
		$p_\nu$ 	&0.160	 			&0.281				&0.523\\
		\hline
		\hline
	\end{tabular}
\end{table}

\begin{figure}[ht!]
	\centering 
\includegraphics[width=\linewidth]{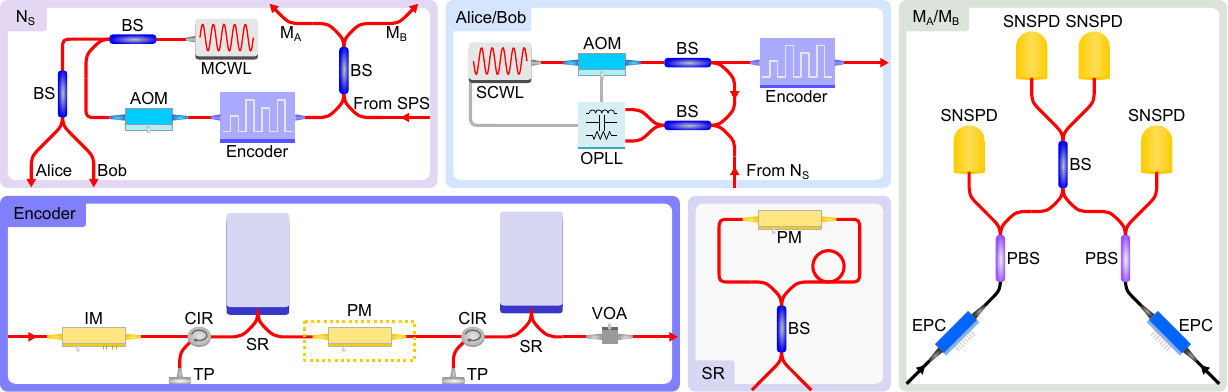}
\caption{Details of the experimental setup. N\textsubscript{S}: an MCWL is located at the central quantum relay node. The continuous-wave light generated by the MCWL is split into two parts: one serves as the frequency reference for the user nodes (Alice and Bob), whereas the other is frequency shifted by $\SI{50}{\MHz}$ using an acousto-optic modulator (AOM). The frequency-shifted part is prepared into phase reference pulses by an encoder. These phase reference pulses and single photons from the single-photon source (SPS) enter the BS in a time-division-multiplexed manner and are then divided into two beams to be sent to measurement nodes M\textsubscript{A} and M\textsubscript{B}. Alice/Bob: settings for the two users. Alice and Bob utilize the same coherent light source and encoder configuration for coherent state modulation, which includes the preparation of phase reference pulses and signal pulses. They use heterodyne optical phase-locked loop (OPLL) technology to lock their SCWLs onto the frequency reference light sent from the central quantum relay node (N\textsubscript{S}). An AOM is inserted to achieve high-speed and wide-range frequency feedback in conjunction with the SCWL. M\textsubscript{A}/M\textsubscript{B}: settings for the two measurement nodes. Before the interference measurement, single photons and coherent pulses require polarization alignment via feedback control, utilizing an electric polarization controller (EPC), a PBS and an SNSPD channel to achieve optimal interference. Encoder: a combination device for the generation and intensity and phase modulation of coherent light pulses. The encoder is composed of an intensity modulator (IM), two Sagnac rings (SRs), a phase modulator (PM) and a variable optical attenuator (VOA). The SR consists of a BS and a PM positioned asymmetrically to achieve different phase modulations for the clockwise and counterclockwise pulses, thereby facilitating intensity modulation. TP, test port.}	
\label{fig:setup}
\end{figure}

\section{Discussion and Outlook}

Our work demonstrates a quantum relay platform utilizing a quantum dot single-photon source, leveraging its unique optical properties to integrate into quantum networks. This design significantly enhances information transmission capacity, coverage, and deployment flexibility while introducing a modular and scalable relay architecture. Our experimental results highlight the suitability of quantum dot single-photon sources as effective relays for information transfer within quantum networks.

Additionally, given that the QKD scheme inherently involves the distribution and distillation of entangled pairs, successfully executing the QKD protocol and distributing keys between users implies the ability of implementing an entanglement distillation protocol, thereby linking distributed quantum systems the quantum computers with our relay structure. Therefore, our results also suggest that quantum dot single-photon sources could serve as relays for entanglement distribution between quantum systems, such as atomic ensembles and ion trap systems. One feasible approach involves the two users each preparing photons entangled with their respective local systems, such as ion-trap and cold-atom quantum memories, and transmitting these photons to the measurement nodes. By performing post-selection based on the measurement results, the local systems of the two users can become entangled. This approach is particularly promising, as entanglement between two atomic ensembles has already been demonstrated through photon transmission over city-scale optical fibers \cite{Yu2020Entanglement}. We believe this method can be directly applied to our relay structure, as our setup fundamentally replaces the interference measurement used in \cite{Yu2020Entanglement} with a single-photon source and two measurement nodes.

Moreover, our relay structure is not restricted to inputs of coherent light from end users. Its design relies fundamentally on photon indistinguishability and interference, and is thus compatible with any light source capable of interference. This includes, for example, quantum emitters that may produce single photons, provided that their spectral and temporal properties are appropriately matched. This flexibility highlights the potential versatility of our rely structure and opens up exciting new avenues for exploration in quantum physics. It is worth noting that integrating photons from different types of sources, such as quantum dots and atomic systems, may require technical developments in spectral reshaping, temporal synchronization, and precise control of photon emission to ensure high interference visibility.

The introduction of quantum dot single-photon source relays provides a critical advantage unique to our structure: the mitigation of dark counts, a major limitation in communication distance. In traditional non-relay communication, as fiber length increases, the probability of a detector response decreases. When this probability becomes comparable to the dark count rate, errors from dark counts dominate, halting successful key generation. Our relay structure inherently halves the distance traveled by individual pulses, improving the signal-to-noise ratio and delaying the point where dark counts dominate the error rate. Our simulation results show that, under ideal conditions, our system achieves the same key rate scaling as twin-field QKD. Moreover, while twin-field QKD fails to generate secure keys beyond approximately $\SI{600}{ \km}$ due to excessive errors caused by detector dark counts, our system is able to distribute secret keys or entanglement over distances exceeding $\SI{1000}{ \km}$ when the intrinsic error rate of the protocol is reduced to around 4–6\%. Detailed calculations and simulation results are provided in the Appendix \ref{sc:sim}. This demonstrates the unique capability of our quantum relay structure to extend quantum communication distances effectively.

Several promising directions can further extend the relay distances achievable with our system. First, interference visibility is closely linked to the protocol's error rate, with high visibility becoming increasingly challenging to maintain over long distances. Enhancing spectral overlap between coherent pulses and single-photon pulses through finer wavelength tuning, faster pulse shaping, and advanced spectral filtering will be critical. Second, simulations show that asymmetrically positioning the measurement nodes relative to the user nodes and single-photon source can improve system performance.  Third, due to the quantum dot’s pulsed laser lacking external clock synchronization, the system’s main clock relies on periodic signals generated by this pulse, introducing instability and increasing error rates. Developing new synchronization schemes will be key to overcoming this challenge. Deploying this relay structure in real-world networks will also require advanced filtering techniques and precise modulation to maintain high interference visibility and minimize errors. These challenges represent crucial areas for future investigation.

Furthermore, the scalability potential of our relay structure is significant. By incorporating additional measurement nodes and single-photon sources, the configuration can be expanded to form sequences such as user node – measurement node – single-photon source – measurement node – single-photon source – measurement node – user node. When the intermediate measurement node between single-photon sources clicks, the photons emitted toward the user nodes effectively mimic the single-photon source in the five-node structure. This concept can be extended to include more single-photon sources and measurement nodes, creating larger relay networks. However, achieving this scalability will require advances in fabricating multiple identical single-photon sources and in coordinating interference across an expanded network. These developments present exciting opportunities for future research.
%%%%%%%%%%%%%%%%%%%%%%%%%%%%%%%%%%%%%%%%
\section{Methods}
Here, we present the detailed experimental setup, as shown in Fig.~\ref{fig:setup}. Below, we will explain the functions of some components in more detail, focusing primarily on those related to phase locking and compensation.

In practice, the frequency difference between Alice's and Bob's lasers, along with phase drift caused by fiber disturbances, can affect the interference results in Eq.~\eqref{eq:phasetran}, introducing additional interference terms. Thus, we modify Eq.~\eqref{eq:phasetran} to the following form:
\begin{equation}
	\sqrt{\mu\eta_1\eta_2}(e^{i(\varphi_A+\Delta \theta)}+e^{i\varphi_B})(a_{3}^\dagger b_{3}^\dagger + a_{4}^\dagger b_{4}^\dagger)\ket{0} + \sqrt{\mu\eta_1\eta_2}(e^{i(\varphi_A+\Delta \theta)}-e^{i\varphi_B})(a_{3}^\dagger b_{4}^\dagger + a_{4}^\dagger b_{3}^\dagger)\ket{0},
\end{equation}
where $\varphi_A$ ($\varphi_B$) represents the encoded phase of Alice (Bob) and $\mu=\abs{\alpha}^2=\abs{\beta}^2$ denotes the intensity. The term $\Delta \theta$ is the phase difference between Alice's and Bob's pulses caused by the instruments, which can be expressed as follows:
\begin{equation}\label{eq:phasediff}
	\Delta \theta=\theta_A+\Delta \omega t+\theta_1-\theta_2-\theta_B+\theta_3-\theta_4,
\end{equation}
where $\theta_A$ ($\theta_B$) is the initial phase of Alice (Bob), $\theta_i$ for $i=1,\dots,4$ represents the phase drift induced by the four fiber links, and $\Delta \omega$ is the angular frequency difference between the two lasers, which causes the phase of the interfering pulses to vary with time $t$.

To ensure that the interference results align closely with the theoretical outcomes and to minimize the additional error rate, it is crucial to reduce $\Delta \omega$ as much as possible. To achieve this, we first establish a stable phase reference between Alice and Bob to minimize $\Delta \omega$. We place a master continuous-wave laser (MCWL) at the single-photon relay node. The MCWL is split into two beams of light, which serve as frequency reference lights for Alice and Bob. Both Alice and Bob employ a heterodyne optical phase-locked loop technique to lock their slave continuous-wave lasers (SCWL) to the frequency reference light, maintaining a stable frequency difference of $\SI{50}{\MHz}$. Additionally, to enhance feedback performance, we insert an acousto-optic modulator to increase the speed of frequency feedback.

To minimize the impact of phase drift, represented by the $\theta$ terms in Eq.~\eqref{eq:phasediff}, we adopted a method that involves sending phase reference pulses, measuring them to estimate the phase drift, and then applying compensation accordingly. In order to prepare and measure phase reference pulses, Alice and Bob utilize their encoders to not only generate signal pulses --- comprising signal, decoy, and vacuum states --- but also phase reference pulses. Each encoder at the user nodes consists of an intensity modulator, two Sagnac rings, a phase modulator, and a variable optical attenuator. By passing through the intensity modulator, the continuous wave generated by the SCWL is converted into light pulses with a pulse width of $\SI{200}{\ps}$ and a repetition rate of $\SI{304.52}{\MHz}$. The Sagnac rings are used for intensity modulation to generate the required phase reference pulses and signal pulses with a $\SI{200}{\MHz}$ repetition rate, and the variable optical attenuator attenuates the pulses down to the single-photon level.

At the single-photon quantum relay node, the remaining continuous-wave light from the MCWL needs to be frequency-shifted by $\SI{50}{\MHz}$ before preparing phase reference pulses using an encoder. Here, the encoder does not require a phase modulator for phase encoding. The prepared phase reference pulses share a beam splitter with the single photons in a time-division multiplexed manner. In this way, to match the coherent state prepared by Alice and Bob, the repetition rate of the single photon also becomes $\SI{200}{\MHz}$. 

Once the pulses prepared by Alice and Bob, as well as the single photons prepared at the quantum relay node, arrive at the measurement nodes, their polarization is corrected by a polarization feedback process. This process is realized by an electric polarization controller, a polarization beam splitter, and one channel of superconducting nanowire single-photon detector (SNSPD). Finally, the polarization-aligned coherent light pulses and single photons enter the beam splitter for interference, whose results are detected by two channels of SNSPD at each measurement node.

\section*{Acknowledgements}

This work was supported by the Innovation Program for Quantum Science and Technology (Grant No. 2021ZD0300702 to T.-Y.C, 2021ZD0301300-1 to T.-Y.C, 2021ZD0300804 to X.M., 2021ZD0301400 to Y.-M.H., 2021ZD0300204 to Y.-H.H, 2021ZD0300800 to Q.Z., X.-P.X., and M.-Y.Z.), the National Natural Science Foundation of China (Grant No. 12174216 to X.M., 12022402 to Y.-M.H, 62474168 to Y.-H.H), the Anhui Initiative in Quantum Information Technologies (Grant No. AHY060000 to C.-Y.L.), the Shanghai Municipal Science and Technology Major Project (Grant No. 2019SHZDZX01 to C.-Y.L. and Y.-H.H), the Chinese Academy of Sciences Project for Young Scientists in Basic Research (Grant No. YSBR-112 to Y.-H.H), the Youth Innovation Promotion Association of CAS (Grant No. Y2023128 to Y.-M.H.), the Independent Deployment Project of HFNL (Grant No. ZB2025010300 to Y.-M.H.), the Strategic Priority Research Program of Chinese Academy of Sciences (Grant No. XDA0520401 to Y.-M.H.), the China Postdoctoral Science Foundation (Grant No. 2022T150628 to X.D.), and the Postdoctoral Research Project in Anhui Province (Grant No. 2024C890 to X.Y.). C.-Y.L. acknowledges support from the New Cornerstone Science Foundation. T.-Y.C. acknowledges support from the Anhui Initiative in Quantum Information Technologies.

\section*{Author contributions}
T.-Y.C., X.M., and J.-W.P. conceived the research. M.Z., Y.-M.H., T.-Y.C., Y.H., X.M., C.-Y.L., and J.-W.P. designed the experiment. M.Z. and Y.-M.H. carried out the experiment and performed data post-processing. Y.H., X.M., performed the protocol security analysis and data post-processing. Y.H., Y.-M.H., and B.-C.L. performed a theoretical interference analysis. J.-Y.Z., R.-Z.L., and Y.-H.H. grew and fabricated the quantum dot samples. Y.-P.G., X.D., G.-Y.Z., H.W., and C.-Y.L.contributed to the generation of single-photon sources. M.-C.X. and Z.N. fabricated the cavity mirror. X.Y., M.-Y.Z., X.-P.X., and Q.Z. contributed to fluorescence upconversion. W.-X.P., and H.-T.Z. assisted with the experiment scheme discussion. D.Q. and X.J. developed the PI circuit board and maintained it for frequency locking. Y.H., X.M., M.Z., Y.-M.H., T.-Y.C., C.-Y.L., and J.-W.P. co-wrote the manuscript, with input from the other authors. All authors discussed the results and proofread the manuscript. J.-W.P. supervised the project.

\section*{Competing interests}
The authors declare no competing interests.

\section*{Data availability}
All data are available from the corresponding authors upon reasonable request. Source Data are provided with this paper.

\section*{Code availability}
The code for simulating the key rate is available from the corresponding authors upon reasonable request.

\appendix

\section{Protocol and Interference} \label{sc:protocol}
\subsection{Quantum Key Distribution Protocol}
We adapt the phase-matching quantum key distribution (QKD) scheme \cite{Ma2018phase} for our five-node quantum link. In this section, we provide a detailed description of the procedure for the phase-matching scheme using the decoy-state method \cite{hwang2003decoy,Lo2005Decoy,wang2005decoy}. The steps are outlined in Box \ref{box:PMQKD}.

\begin{mybox}[label={box:PMQKD}]{Phase-matching QKD scheme with a single-photon relay}
\begin{enumerate}[(1)]
\item 
State Preparation: In the $i$-th round, Alice prepares a coherent state $\ket{\alpha_i}=\ket{\sqrt{\mu_a^i}e^{\mathrm{i} (\pi\kappa^i_a+\phi^i_a)}}_{a_1}$ on optical mode $a_1$, where the intensity $\mu_a^i$ is randomly chosen from $\{0, \nu, \mu\}$, $\kappa^i_a$ is Alice's raw key bit chosen randomly from $\{0, 1\}$, and phase $\phi^i_a$ is chosen uniformly from $D=16$ discrete values $\{0, \pi/8, \dots, 15\pi/8\}$. The intensities satisfy $0 < \nu < \mu$. Similarly, Bob randomly chooses $\mu_b^i$, $\kappa^i_b$, and $\phi^i_b$, and prepares $\ket{\beta_i}=\ket{\sqrt{\mu_b^i}e^{\mathrm{i} (\pi\kappa^i_b+\phi^i_b)}}_{b_1}$ on mode $b_1$. At the relay node, a single-photon pulse passes through a 50:50 beam splitter, with the outgoing paths denoted as optical modes $a_2$ and $b_2$.

\item
Measurement: Alice and Bob send their optical modes $a_1$ and $b_1$ to the untrusted measurement nodes $C_a$ and $C_b$, respectively. Simultaneously, the optical modes $a_2$ and $b_2$ are sent to $C_a$ and $C_b$, respectively. The measurement nodes $C_a$ and $C_b$ perform interference and record clicks on detectors $L$ and/or $R$. A valid detection occurs when exactly one detector clicks at a measurement node.

\item %\label{StepAnnounce}
Announcement: For all rounds with valid detection, the two measurement nodes announce the detection results ($L$ or $R$). If both detection nodes announce valid detection within the same round, Alice and Bob consider that round as successful. Alice and Bob then announce their random phases, $\phi^i_a$ and $\phi_b^i$, for all successful rounds.

\item %\label{StepParaEst}
Key Mapping: Alice and Bob repeat the above steps for $N$ rounds. For successful rounds, they compare their encoded random phases $\phi^i_a$ and $\phi^i_b$. If $\abs{\phi^i_a - \phi^i_b} = 0$ or $\pi$, Alice and Bob keep their respective raw key bits $\kappa_a^i$ and $\kappa_b^i$ for that round. Otherwise, they discard their raw key bits. Additionally, Bob flips his key bit $\kappa_b^i$ if the announcements from the two measurement nodes differ or if $\abs{\phi^i_a - \phi^i_b} = \pi$.

\item %\label{InfoRecon}
Information Reconciliation: Alice and Bob reconcile their raw key bits over a public channel. Then, they perform error verification to ensure successful error correction, record the amount of classical communication used during this process, and count the number of errors.

\item %\label{StepKeymeasure}
Parameter Estimation: Alice and Bob assess the parameters for all remaining raw key bits, focusing on the gains $Q_\mu$ and the quantum bit error rates $E_Z$. Here, the gain $Q_\mu$ represents the probability of successful detection for the signal states with intensity $\mu$, while the quantum bit error rates $E_Z$ are determined after information reconciliation. Subsequently, they estimate the phase error rate $e_p$ following the methodology in \cite{Zeng2019Symmetryprotected}.

\item %\label{StepClassical}
Privacy Amplification: Alice and Bob evaluate the privacy amplification ratio based on the key rate. They then apply privacy amplification to produce the final secure key bits.
\end{enumerate}
\end{mybox}

Note that the phase-matching scheme employed here follows the original scheme in \cite{Ma2018phase}, with the main difference being the addition of a single-photon source relay and two detection nodes in the system. In essence, as long as the results from the two detection nodes play the same role as the results from a single detection node in the original scheme, Alice and Bob follow the same steps as in the original. In security analysis, all three intermediate nodes can be treated as one untrusted Bell-state measurement. This is why the previous security proofs based on symmetry-protected privacy \cite{Zeng2019Symmetryprotected} and the source-replacement model \cite{huang2023source} are directly applicable to our system.

\subsection{Interference Analysis}
We now proceed to analyze why this five-node structure can effectively reflect the phase difference between user-encoded quantum states through interference results. This relates to the statement mentioned in the main text: \emph{"When the phases of \(\ket{\alpha}\) and \(\ket{\beta}\) are the same, the detectors on the same side of both measurement sites will click simultaneously. When the phase difference is \(\pi\), the detectors on opposite sides will click."}

\begin{figure}
	\centering \includegraphics[width=12cm]{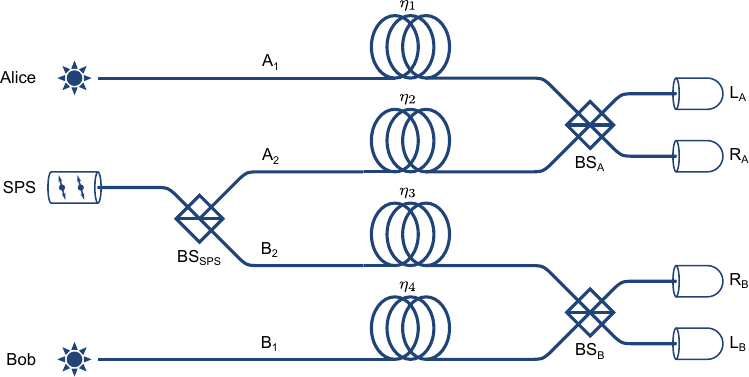}
	\caption{A schematic of the five-node structure. This figure corresponds to Figure 1 in the main text. For clarity, only the light source, optical fibers, beam splitters, and detectors from the experimental setup are depicted. The transmittance of the four paths is denoted as $\eta_i, i=1,2,3,4$. BS: beam splitters; SPS: single-photon source.}\label{fig:setting2}
\end{figure}

To clarify the description of the quantum states at different stages in the protocol, we simplify the experimental setup into the form illustrated in Fig.~\ref{fig:setting2}.

Firstly, we demonstrate how Alice and Bob can determine the phase difference between their quantum states based on the click results of the detectors. The state on modes $A_1$ and $B_1$ can be represented by the creation operators as:
\begin{equation}
	\begin{split}
		\ket{\alpha} &= e^{-\abs{\alpha}^2/2}e^{\alpha a_1^\dagger}\ket{0}, \\
		\ket{\beta} &= e^{-\abs{\beta}^2/2}e^{\beta b_1^\dagger}\ket{0}.
	\end{split}
\end{equation}
Here, lowercase letters and subscripts denote the creation operator on the corresponding modes. For instance, $a_1^\dagger$ represents the creation operator on mode $A_1$. The state on modes $A_2$ and $B_2$, generated from a single photon, is given by
\begin{equation}
	\dfrac{1}{\sqrt{2}}\left(a_2^\dagger+b_2^\dagger\right)\ket{0}=\dfrac{1}{\sqrt{2}}\left(\ket{10}+\ket{01}\right).
\end{equation}

After passing through the optical fibers, the intensities of the coherent states are attenuated as follows:
\begin{equation}
	\begin{split}
		\alpha &\rightarrow \sqrt{\eta_1}\alpha \equiv \alpha',\\
		\beta &\rightarrow \sqrt{\eta_4}\beta \equiv \beta'.
	\end{split}
\end{equation}
As for the single-photon state, it effectively passes through a beam splitter with transmittances $\eta_2$ and $\eta_3$. Thus, the resulting state on modes $A_2$ and $B_2$ can be expressed as
\begin{equation}
	\begin{split}
		\left(\sqrt{\frac{\eta_2}{2}}a_2^\dagger+\sqrt{\frac{\eta_3}{2}}b_2^\dagger + \sqrt{\frac{1-\eta_2}{2}}\tilde{a}_2^\dagger + \sqrt{\frac{1-\eta_3}{2}}\tilde{b}_2^\dagger\right) \ket{0},
	\end{split}
\end{equation}
where $\tilde{a}_2^\dagger$ and $\tilde{b}_2^\dagger$ denote the loss components in modes $A_2$ and $B_2$, respectively. Since we are not concerned with the loss of the single photon, we coarse-grain these two operators and define 
\begin{equation}
	c^\dagger = \frac{\sqrt{\frac{1-\eta_2}{2}}\tilde{a}_2^\dagger + \sqrt{\frac{1-\eta_3}{2}}\tilde{b}_2^\dagger}{\sqrt{1-\frac{\eta_2+\eta_3}{2}}}
\end{equation}
as the overall loss operator for the single photon. Thus, the state before entering $\text{BS}_a$ and $\text{BS}_b$ is given by:
\begin{equation}\label{eq:inputstate}
\begin{split}
e^{-(\eta_1\abs{\alpha}^2+\eta_4\abs{\beta}^2)/2}e^{\sqrt{\eta_1}\alpha a_1^\dagger}e^{\sqrt{\eta_4}\beta b_1^\dagger} \left(\sqrt{\frac{\eta_2}2}a_2^\dagger+\sqrt{\frac{\eta_3}2}b_2^\dagger+\sqrt{1-\frac{\eta_2+\eta_3}2}c^\dagger\right) \ket{0} \\
\end{split}
\end{equation}

Assuming lossy channels with $\eta_1\abs{\alpha}^2 \ll 1$ and $\eta_4\abs{\beta}^2 \ll 1$, we can expand the $\mathrm{e}$ exponent, ignore the higher-order terms, and retain only up to the second-order terms in Eq.~\eqref{eq:inputstate}:
\begin{equation}\label{eq:BSinputApp}
	\begin{split}
		\left[1+\sqrt{\eta_1}\alpha a_1^\dagger + \sqrt{\eta_4}\beta b_1^\dagger + \sqrt{\eta_1\eta_4}\alpha\beta a_1^\dagger b_1^\dagger + \frac{1}{2} \eta_1 \alpha^2 (a_1^\dagger)^2 + \frac{1}{2} \eta_4 \beta^2 (b_1^\dagger)^2\right] 
		\left(\sqrt{\eta_2}a_2^\dagger + \sqrt{\eta_3}b_2^\dagger + \sqrt{2-\eta_2-\eta_3} c^\dagger \right) \ket{0},
	\end{split}
\end{equation}
where the state is not normalized in general, and we consider the normalization factor at the end. In the detection post-selection, Alice and Bob only keep events where there is one click on either $L_A$ or $R_A$, and one click on either $L_B$ or $R_B$. Thus, there must be at least one photon in modes $A_1$ or $B_1$. After post-selection, the state becomes:
\begin{equation}\label{eq:BSinputPost}
	\begin{split}
		&\left(\sqrt{\eta_1\eta_3}\alpha a_1^\dagger b_2^\dagger + \sqrt{\eta_2\eta_4}\beta a_2^\dagger b_1^\dagger + \sqrt{\eta_1\eta_4(2-\eta_2-\eta_3)}\alpha\beta a_1^\dagger b_1^\dagger c^\dagger \right) \ket{0} \\
		&+ \left(\sqrt{\eta_1\eta_4}\alpha\beta a_1^\dagger b_1^\dagger \sqrt{\eta_2}a_2^\dagger 
		+ \frac{1}{2} \eta_4 \beta^2 (b_1^\dagger)^2 \sqrt{\eta_2}a_2^\dagger
		+ \sqrt{\eta_1\eta_4}\alpha\beta a_1^\dagger b_1^\dagger \sqrt{\eta_3}b_2^\dagger
		+ \frac{1}{2} \eta_1 \alpha^2 (a_1^\dagger)^2 \sqrt{\eta_3}b_2^\dagger\right) \ket{0}.
	\end{split}
\end{equation}
Next, we further ignore the terms with orders of $\eta_i$ higher than $\frac{3}{2}$ and only keep the first order of $\eta_i$:
\begin{equation}\label{eq:BSinputNohighOrder}
	\begin{split}
		\left(\sqrt{\eta_1\eta_3}\alpha a_1^\dagger b_2^\dagger + \sqrt{\eta_2\eta_4}\beta a_2^\dagger b_1^\dagger + \sqrt{\eta_1\eta_4(2-\eta_2-\eta_3)}\alpha\beta a_1^\dagger b_1^\dagger c^\dagger \right) \ket{0}. \\
	\end{split}
\end{equation}

Then, the single photon and the coherent states interfere at the beam splitters:
\begin{equation}\label{eq:BS}
	\begin{split}
		&a_1^\dagger \stackrel{\text{BS}_a}\longrightarrow\dfrac{1}{\sqrt{2}}(a_{3}^\dagger+a_{4}^\dagger),\\
		&a_2^\dagger \stackrel{\text{BS}_a}\longrightarrow\dfrac{1}{\sqrt{2}}(a_{3}^\dagger-a_{4}^\dagger),\\
		&b_1^\dagger \stackrel{\text{BS}_b}\longrightarrow\dfrac{1}{\sqrt{2}}(b_{3}^\dagger-b_{4}^\dagger),\\
		&b_2^\dagger \stackrel{\text{BS}_b}\longrightarrow\dfrac{1}{\sqrt{2}}(b_{3}^\dagger+b_{4}^\dagger).\\
	\end{split}
\end{equation}
The final state detected by the single-photon detectors can be expressed as:
\begin{equation}\label{eq:statedetected}
	\begin{split}
		&\left(\sqrt{\eta_1\eta_3}\alpha (a_{3}^\dagger + a_{4}^\dagger) (b_{3}^\dagger + b_{4}^\dagger) + \sqrt{\eta_2\eta_4}\beta (a_{3}^\dagger - a_{4}^\dagger) (b_{3}^\dagger - b_{4}^\dagger)\right) \ket{0} \\
		&\quad + \sqrt{\eta_1\eta_4(2-\eta_2-\eta_3)}\alpha\beta (a_{3}^\dagger + a_{4}^\dagger) (b_{3}^\dagger - b_{4}^\dagger) c^\dagger \ket{0} \\
		&= (\sqrt{\eta_1\eta_3}\alpha + \sqrt{\eta_2\eta_4}\beta)(a_{3}^\dagger b_{3}^\dagger + a_{4}^\dagger b_{4}^\dagger)\ket{0} \\
		&\quad + (\sqrt{\eta_1\eta_3}\alpha - \sqrt{\eta_2\eta_4}\beta)(a_{3}^\dagger b_{4}^\dagger + a_{4}^\dagger b_{3}^\dagger)\ket{0} \\
		&\quad + \sqrt{\eta_1\eta_4(2-\eta_2-\eta_3)}\alpha\beta (a_{3}^\dagger b_{3}^\dagger - a_{4}^\dagger b_{4}^\dagger - a_{3}^\dagger b_{4}^\dagger + a_{4}^\dagger b_{3}^\dagger) c^\dagger \ket{0}.
	\end{split}
\end{equation}
Note that the three terms are orthogonal to each other. The first two terms are in different modes, while the third term contains the environment mode with the creation operator $c^\dagger$.

The first two terms in Eq.~\eqref{eq:statedetected} represent cases where one click is caused by the single photon, and the other is caused by the coherent state. As shown below, these two terms reflect the phase difference between the two coherent states, which is crucial for key generation. The third term represents a situation where the single photon is lost in the transmission channel and both clicks are caused by the coherent states. In this case, the click results are random, leading to a 50\% error rate during key generation.

To achieve perfect interference of the first two terms in Eq.~\eqref{eq:BSinputPost}, Alice and Bob need to set their intensities to satisfy:
\begin{equation}\label{eq:InterfereCond}
	\abs{\sqrt{\eta_1\eta_3}\alpha} = \abs{\sqrt{\eta_2\eta_4}\beta}.
\end{equation}

If $\phi_a = \phi_b$, the second term in Eq.~\eqref{eq:statedetected} cancels, and the state becomes:
\begin{equation}\label{eq:phia=phib}
	(\sqrt{\eta_1\eta_3}\alpha + \sqrt{\eta_2\eta_4}\beta)(a_{3}^\dagger b_{3}^\dagger + a_{4}^\dagger b_{4}^\dagger)\ket{0} 
	+ \sqrt{\eta_1\eta_4(2-\eta_2-\eta_3)}\alpha\beta (a_{3}^\dagger b_{3}^\dagger - a_{4}^\dagger b_{4}^\dagger - a_{3}^\dagger b_{4}^\dagger + a_{4}^\dagger b_{3}^\dagger)c^\dagger \ket{0}.
\end{equation}
The first term in Eq.~\eqref{eq:phia=phib} results in click events $L_a\bar{R}_a L_b\bar{R}_b$ or $\bar{L}_a R_a \bar{L}_b R_b$, allowing Alice and Bob to generate correct raw key bits. For the second term, there is a 50\% probability of obtaining click events $L_a\bar{R}_a L_b\bar{R}_b$ or $\bar{L}_a R_a \bar{L}_b R_b$, while the other 50\% results in $L_a\bar{R}_a \bar{L}_b R_b$ or $\bar{L}_a R_a L_b\bar{R}_b$, causing a key bit error between Alice and Bob. Therefore, the error rate is:
\begin{equation}\label{eq:error0}
	e = \frac{\eta_1\eta_4(2-\eta_2-\eta_3)\abs{\alpha\beta}^2}{\abs{\sqrt{\eta_1\eta_3}\alpha + \sqrt{\eta_2\eta_4}\beta}^2 + 2\eta_1\eta_4(2-\eta_2-\eta_3)\abs{\alpha\beta}^2}.
\end{equation}

Similarly, if $\phi_a = \phi_b + \pi$, the first term in Eq.~\eqref{eq:statedetected} cancels, and the state becomes:
\begin{equation}\label{eq:phia=phibpi}
	(\sqrt{\eta_1\eta_3}\alpha - \sqrt{\eta_2\eta_4}\beta)(a_{3}^\dagger b_{4}^\dagger + a_{4}^\dagger b_{3}^\dagger)\ket{0}
	+ \sqrt{\eta_1\eta_4(2-\eta_2-\eta_3)}\alpha\beta (a_{3}^\dagger b_{3}^\dagger - a_{4}^\dagger b_{4}^\dagger - a_{3}^\dagger b_{4}^\dagger + a_{4}^\dagger b_{3}^\dagger)c^\dagger \ket{0}.
\end{equation}
The first term in Eq.~\eqref{eq:phia=phibpi} results in click events $L_a\bar{R}_a \bar{L}_b R_b$ or $\bar{L}_a R_a L_b \bar{R}_b$, which correspond to correct raw key bits. The second term still has a 50\% probability of causing a key bit error. Thus, the error rate is:
\begin{equation}\label{eq:errorpi}
	e = \frac{\eta_1\eta_4(2-\eta_2-\eta_3)\abs{\alpha\beta}^2}{\abs{\sqrt{\eta_1\eta_3}\alpha - \sqrt{\eta_2\eta_4}\beta}^2 + 2\eta_1\eta_4(2-\eta_2-\eta_3)\abs{\alpha\beta}^2}.
\end{equation}
Considering Eq.~\eqref{eq:InterfereCond} and the phase relationship between $\alpha$ and $\beta$, the error rates in Eqs.~\eqref{eq:error0} and \eqref{eq:errorpi} are the same and are given by:
\begin{equation}
	e = \frac{\eta_4(2-\eta_2-\eta_3)\abs{\beta}^2}{4\eta_3 + 2\eta_4(2-\eta_2-\eta_3)\abs{\beta}^2}.
\end{equation}

In a symmetric setting where $\eta_1 = \eta_4$, $\eta_2 = \eta_3$, and $\abs{\alpha} = \abs{\beta} = \sqrt{\mu}$, the error rate simplifies to:
\begin{equation}
	e = \frac{\eta_1(1-\eta_2)\mu}{2\eta_2 + \eta_1(1-\eta_2)\mu}.
\end{equation}
From this, it is evident that $\mu$ must be kept small in order to suppress errors effectively.

\subsection{Theoretical model of single-photon and coherent-state interference}
The above analysis is based on the premise of coincident detection at the two measurement nodes. In our experiment, analyzing the interference results between the coherent state and the single photon within a single measurement node is also crucial. Therefore, in this subsection, we will similarly provide the derivation of the interference results. Considering that in practical experiments, interference visibility is further influenced by factors such as spectral overlap, it cannot be simply summarized using the creation and annihilation operator model described in the previous section. Therefore, in this subsection, we will provide a more detailed derivation starting from the cross-correlation function.

\begin{figure}
\centering \includegraphics[width=0.5\linewidth]{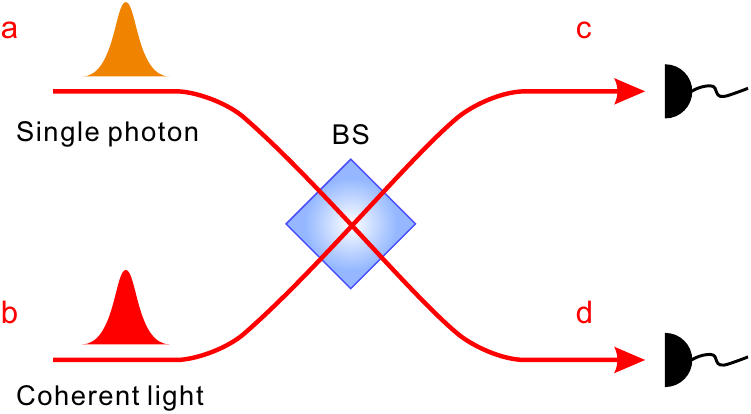}
\caption{Schematic of the interference between coherent light pulses and single photons.}\label{fig:SPSCoh}
\end{figure}

The basic interference setup is shown in Fig.~\ref{fig:SPSCoh}. As discussed in the main text, the single photon and the weak coherent laser interfere after traveling through several tens of kilometers of fiber fiber and reaching a 50/50 beamsplitter. In this setup, both the single photon and the weak coherent laser are extremely weak. Here, both the single-photon state and the weak coherent state can be described as pure states. Additionally, since the experiment is conducted under low-intensity conditions, we neglect multi-photon events involving more than two photons.

%Furthermore, any pure quantum state can be represented as a linear combination in the Fock state basis.

We denote the probability of emitting vacuum, one photon, and two photons from the quantum dot (QD) within one pulse period as $p_0$, $p_1$, and $p_2$, respectively. 
Then, the state emitted from QD in the Fock state basis can be written as 
\begin{equation}
	\ket{\psi_{\text{QD}}}=\sqrt{p_0}\ket{0_a}+\sqrt{p_1}\ket{1_a}+\sqrt{p_2}\ket{2_a}.
\end{equation}

Similarly, we denote the probability of vacuum, one photon, and two photons emitted from the laser within one pulse period as $\tilde{p}_0$, $\tilde{p}_1$ and $\tilde{p}_2$. Then, the state from the laser is 
\begin{equation}
	\ket{\psi_{\text{laser}}}=\sqrt{\tilde{p}_0}\ket{0_a}+\sqrt{\tilde{p}_1}\ket{1_a}+\sqrt{\tilde{p}_2}\ket{2_a}.
\end{equation}

In the low-efficiency regime, the probability of the vacuum state component in both the single-photon state and the coherent state is very high, approaching 1. Consequently, the following approximation can be made:
\begin{equation}
p_0 \rightarrow 1, \quad \tilde{p}_0 \rightarrow 1.
\end{equation}

With these states, we can calculate the second-order correlation function of $g^{2}(0)$ as follows:
\begin{equation}
g^{2}(0)=\dfrac{\bra{\psi}(a^\dagger)^2 a^2\ket{\psi}}{\abs{\bra{\psi}a^\dagger a\ket{\psi}}^2},
\end{equation}
Where $a^\dagger$ and $a$ are creation and annihilation operator on the optical mode, respectively. Then, the $g^{2}(0)$ for the two states are
\begin{equation}
\begin{split}
g^{2}_\text{QD}(0) &= \dfrac{2p_2}{\left(p_1\right)^2},\\
g^{2}_\text{laser}(0) &= \dfrac{2\tilde{p}_2}{\left(\tilde{p}_1\right)^2}=1.
\end{split}
\end{equation}
Here $g^{2}_\text{laser}(0)=1$ since it is coherent state. Thus, these two states can be approximated as
\begin{equation}
\begin{split}
	\ket{\tilde{\psi}_{\text{QD}}}&=A_1\left(\ket{0_a}+\sqrt{p_1}\ket{1_a}+\sqrt{\dfrac{g^{2}_\text{QD}(0)(p_1)^2}{2}}\ket{2_a}\right), \\
	p_2 &= \dfrac{g^{2}_\text{QD}(0)(p_1)^2}{2},\\
	\ket{\tilde{\psi}_{\text{laser}}}&=A_2\left(\ket{0_a}+\sqrt{\tilde{p}_1}\ket{1_a}+\sqrt{0.5(\tilde{p}_1)^2}\ket{2_a}\right), \\
	\tilde{p}_2 &= 0.5(\tilde{p}_1)^2,
\end{split}
\end{equation}
where $A_1$ and $A_2$ are the normalization coefficients.

Then, we consider the interference results. The probability of a coincidence event at detector $c$ at time $t$ and at detector $d$ at time $t+\tau$ can be calculated as,
\begin{equation}
P_\text{joint}(\mathrm{c,t;d,t+\tau})=K\left\langle\hat{E}_{c}^{-}(\mathrm{t})\hat{E}_{d}^{-}(\mathrm{t+\tau})\hat{E}_{d}^{+}(\mathrm{t+\tau})\hat{E}_{c}^{+}(\mathrm{t})\right\rangle,
\end{equation}
where $K$ is a constant.

Then, we let $T_c(T_d)$ denote the transit times of photons in the interference setup. The transmission and reflection ratios of the beam splitter are denoted as $T$ and $R$ respectively. We then have \cite{glauber1963quantum}

\begin{equation}
\begin{aligned}
	&\left\langle\hat{E}_{c}^{-}(\mathrm{t})\hat{E}_{d}^{-}(\mathrm{t}+\tau)\hat{E}_{d}^{+}(\mathrm{t}+\tau)\hat{E}_{c}^{+}(\mathrm{t})\right\rangle\\
	=&\frac{1}{2}p_{2}\tilde{p}_{0}K^{\prime2}RT|\zeta_{a1}\left(t+\tau-T_{d}\right)\zeta_{a2}\left(t-T_{c}\right)+\zeta_{a2}\left(t+\tau-T_{d}\right)\zeta_{a1}\left(t-T_{c}\right)|^{2}\\
	&+\frac{1}{2}p_{0}\tilde{p}_{2}K^{\prime2}RT|\zeta_{b1}\left(t+\tau-T_{d}\right)\zeta_{b2}\left(t-T_{c}\right)+\zeta_{b2}\left(t+\tau-T_{d}\right)\zeta_{b1}\left(t-T_{c}\right)|^{2}\\
	&+\frac{1}{2}p_{1}\tilde{p}_{1}K^{\prime2}\left|T\zeta_{a}\left(t+\tau-T_{d}\right)\zeta_{b}\left(t-T_{c}\right)-R\zeta_{a}\left(t-T_{c}\right)\zeta_{b}\left(t+\tau-T_{d}\right)\right|^{2},
\end{aligned}
\end{equation}
where $K^{\prime}$ is a constant calculated from $K$. Here $\zeta_{a1}$ and $\zeta_{a2}$ are the spatial-temporal functions for the quantum dot. $\zeta_{b1}$ and $\zeta_{b2}$ are defined similarly for that of the laser. The first term in the probability represents the contribution from two-photon events from the QD. The second term is from the laser, and the third term accounts for coincidence events from two independent sources, including contributions from interference. Notably, the two photons from the QD are synchronized with the laser pulse with minimal time jitter due to their generation mechanism and temporal broadening through the grating, while the photons from the laser are randomly distributed in time. 
We denote the time jitter for photons from the QD (laser) as $\Delta\tau_{a1}(\Delta\tau_{b1})$ and $\Delta\tau_{a2}(\Delta\tau_{b2}).$ For the ensemble of all emission events, $\Delta\tau_{a1}$ and $\Delta\tau_{a2}$ are distributed in time as Dirac delta functions, while$\Delta\tau_{b1}$ and $\Delta\tau_{b2}$ are uniformly distributed. The probability for a coincident event with a delay is: 
\begin{equation}
	P_\text{joint}\left(\tau\right)=\left\langle P_\text{joint1}\left(0,\Delta\tau_{a1},\Delta\tau_{a2}\right)\right\rangle|_{\Delta\tau_{a1},\Delta\tau_{a2}}+\left\langle P_\text{joint2}\left(0,\Delta\tau_{b1},\Delta\tau_{b2}\right)\right\rangle|_{\Delta\tau_{b1},\Delta\tau_{b2}}+\left\langle P_\text{joint3}\left(0,\Delta\tau_{a},\Delta\tau_{b}\right)\right\rangle|_{\Delta\tau_{a},\Delta\tau_{b}}
\end{equation}
We use a Lorentzian profile for photons from both sources since they have passed through a grating before interference for approximation. The pulse period is $T_p$ and the duration of data collection is $T_0$. The spectral linewidth after passing through the grating is $\sigma_a$. The spatial-temporal distribution time is $T_g$. The derivation of $P_\text{joint}$ can be seen in \cite{Deng2019quantum}. 

The raw interference visibility depends on the relative ratio of the intensities (photon count rates) from the quantum dot and the laser. In the experiment, we optimize this ratio to achieve maximum raw visibility. The dependence of coincidence probability around zero-time delay on the intensity ratio can be calculated as:
\begin{equation}
\begin{split}
	&\left\langle P_\text{joint1}(0,\Delta\tau_{a1},\Delta\tau_{a2})\right\rangle|_{\Delta\tau_{a1},\Delta\tau_{a2}}=2\sigma_{a}p_{2}\tilde{p}_{0}K^{^{\prime}2}RT,\\
	&\left\langle P_\text{joint2}(0,\Delta\tau_{b1},\Delta\tau_{b2})\right\rangle|_{\Delta\tau_{b1},\Delta\tau_{b2}}=2\frac{T_{g}}{T_{p}^{2}}p_{0}\tilde{p}_{2}K^{\prime2}RT,\\
	&\left\langle P_\text{joint3}\left(0,\Delta\tau_{a},\Delta\tau_{b}\right)\right\rangle|_{\Delta\tau_{a},\Delta\tau_{b}}=\frac{1}{T_{p}}p_{1}\tilde{p}_{1}K^{\prime2}\left(T^{2}+R^{2}-2TR\right).\\
\end{split}
\end{equation}
The above derivation holds for the perfectly indistinguishable scenario. Here, we further introduce a new parameter $V_c$, which represents the corrected interference visibility (corrected by subtracting multi-photon events, dark counts, etc.). In the partially distinguishable scenario, it is straightforward to find:
\begin{equation}
	\left\langle P_\text{joint3}(0,\Delta\tau_{a},\Delta\tau_{b},V_{c})\right\rangle|_{\Delta\tau_{a},\Delta\tau_{b}}=\frac{1}{T_{p}}p_{1}\tilde{p}_{1}K^{^{\prime}2}\left(T^{2}+R^{2}-2V_{c}TR\right)
\end{equation}
Thus, the raw interference visibility is:
%\begin{equation}
%	V_{r}=\frac{\frac{1}{T_{p}}p_{1}\tilde{p}_{1}V_{c}}{\frac{g_\text{QD}^{2}(0)p_{1}^{2}}{2}\sigma_{a}+\frac{T_{g}}{T_{p}^{2}}\tilde{p}_{1}^{2}+\frac{1}{T_{p}}p_{1}\tilde{p}_{1}}
%\end{equation}
\begin{equation}
\begin{aligned}
V_{r}&=\frac{\left\langle P_\text{joint3}(0,\Delta\tau_{a},\Delta\tau_{b},0)\right\rangle|_{\Delta\tau_{a},\Delta\tau_{b}}-\left\langle P_\text{joint3}(0,\Delta\tau_{a},\Delta\tau_{b},V_{c})\right\rangle|_{\Delta\tau_{a},\Delta\tau_{b}}}{\left\langle P_\text{joint1}(0,\Delta\tau_{a1},\Delta\tau_{a2})\right\rangle|_{\Delta\tau_{a1},\Delta\tau_{a2}}+\left\langle P_\text{joint2}(0,\Delta\tau_{b1},\Delta\tau_{b2})\right\rangle|_{\Delta\tau_{b1},\Delta\tau_{b2}}+\left\langle P_\text{joint3}(0,\Delta\tau_{a},\Delta\tau_{b},0)\right\rangle|_{\Delta\tau_{a},\Delta\tau_{b}}}\\
&=\frac{\frac{2}{T_{p}}p_{1}\tilde{p}_{1}K^{\prime2}V_{c}TR}{2\sigma_{a}p_{2}K^{\prime2}TR+2\frac{T_{g}}{T_{p}^{2}}\tilde{p}_{2}K^{\prime2}TR+\frac{1}{T_{p}}p_{1}\tilde{p}_{1}K^{\prime2}(T^{2}+R^{2})}\\
&=\frac{\frac{1}{T_{p}}p_{1}\tilde{p}_{1}V_{c}}{\frac{g_\text{QD}^{2}(0)p_1^2}{2}\sigma_{a}+\frac{T_{g}}{T_{p}^{2}}\tilde{p}_{1}^{2}+\frac{1}{T_{p}}p_{1}\tilde{p}_{1}}
\end{aligned}
\end{equation}

After considering the experiment parameters, we have
\begin{equation}
	\begin{aligned}
		&V_{r}=\frac{V_{c}}{\frac{T_{p}g_\text{QD}^{2}(0)\sigma_{a}}{2}\frac{I_\text{QD}}{I_\text{laser}}\frac{T_{g}}{T_{p}}+0.5\frac{T_{g}}{T_{p}}\frac{I_\text{laser}}{I_\text{QD}}\frac{T_{P}}{T_{g}}+1}\\
		&=\frac{V_{c}}{\frac{g_\text{QD}^{2}(0)\sigma_{a}}{2}\frac{I_\text{QD}}{I_\text{laser}}T_{g}+0.5\frac{I_\text{laser}}{I_\text{QD}}+1},
	\end{aligned}
\end{equation}
where $I_\text{QD}$ and $I_\text{laser}$ are the intensities of the single-photon source and laser.
This results are used to give the simulation results in Fig.~3 in the main text.

In our experiment, we reduce $g_\text{QD}^{2}(0)$ by calibrating the cross-polarization technique to eliminate the laser background and employing a narrow 5 GHz grating to filter out the laser noise. Numerical results indicate that $V_r$ is highly sensitive to $g_\text{QD}^{2}(0)$. Therefore, in the experiment, it is crucial to maintain a low $g_\text{QD}^{2}(0)$ through nearly perfect blocking of the laser background.

\section{Simulation Method}\label{sc:sim}
Building on the interference analysis from the previous section, this section establishes a model for the imperfect single-photon source and introduces the simulation steps used to evaluate the performance of the protocol.

\subsection{Model for Single-Photon Source}\label{sc:nonidealsource}
\begin{figure}[hbpt!]
	\centering \includegraphics[width=4cm]{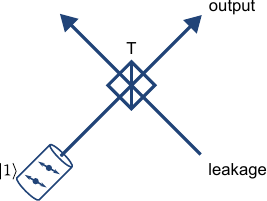}
	\caption{Model of an imperfect single-photon source. Here, we consider the efficiency $T$ of the single-photon source as if a perfect single-photon source passes through a beam splitter with a transmission rate $T$. The non-zero $g^{(2)}(0)$ is attributed to leaked pump light mixing with the output, represented as the ``leakage'' part in the figure.}\label{fig:imperfectsps}
\end{figure}

Here, we model the single-photon source and account for its imperfections. Due to operations like pumping and down-conversion, the state generated by the single-photon source is not perfect. Ideally, a perfect single-photon state can be represented as $\ket{1}$ in the photon number basis. In reality, the quantum dot source does not always emit a single photon but produces a single-photon state with a certain probability. Thus, we model this process as a perfect single-photon state passing through a beam splitter with transmittance $T$, yielding an imperfect single-photon state, as depicted in Fig.~\ref{fig:imperfectsps}. Consequently, the single-photon component in the output pulse is:
\begin{equation}
	\sqrt{T}\ket{1}_s+\sqrt{1-T}\ket{0}_s,
\end{equation}
where the subscript $s$ denotes that the single-photon state is in the signal mode.

In addition to the single-photon state, the output pulse also contains a coherent state due to the leakage of the pumping laser. In practice, the frequency of the single-photon state generated by the quantum dot is approximately 2 to 3 GHz, while the pumping laser pulses have a frequency spectrum distributed over tens of GHz. Therefore, we consider the single-photon state and the leaking coherent state to occupy different modes. Thus, the output state of the quantum dot source is:
\begin{equation}
	\ket{\psi}_{out} = \left(\sqrt{T}\ket{1}_s+\sqrt{1-T}\ket{0}_s\right) \otimes \ket{\gamma}_c,
\end{equation}
where $c$ denotes the mode occupied by the leaking coherent state. The coherent state $\ket{\gamma}_c$ has a form given by $\gamma = \abs{\gamma}e^{\mathrm{i}\theta_\gamma}$, where $\theta_\gamma$ is a completely random phase. The intensity of the leaked coherent state is denoted by $\nu = \abs{\gamma}^2$. The total intensity of the output state is:
\begin{equation}\label{eq:inten}
	I = T + \nu.
\end{equation}

We also have another observable, $g^{(2)}(0)$, that helps characterize the imperfect source. If we apply our model to the Hanbury-Brown-Twiss setting, we have:
\begin{equation}
	\begin{split}
		g^{(2)}(0)&=\dfrac{\langle n_1 n_2 \rangle}{\langle n_1 \rangle \langle n_2 \rangle} \\
		&= \dfrac{\left(\dfrac{\nu}{2} + T\right)\dfrac{\nu}{2}}{\left(\dfrac{T}{2} + \dfrac{\nu}{2}\right)^2}.
	\end{split}
\end{equation}
Consider the scenario where $\nu \ll T$, allowing us to ignore the second-order terms of $\nu$ and simplify the expression to:
\begin{equation}
	g^{(2)}(0) = \dfrac{\nu}{\dfrac{T}{2} + \nu}.
\end{equation}
From this, we can determine the ratio $\dfrac{\nu}{T}$ using $g^{(2)}(0)$:
\begin{equation}\label{eq:g2relation}
	\dfrac{\nu}{T} = \dfrac{\dfrac{1}{2}g^{(2)}(0)}{1-g^{(2)}(0)}.
\end{equation}
Therefore, with the total intensity $I$ and the second-order degree $g^{(2)}(0)$, we can determine the parameters $T$ and $\nu$ to characterize the output state. Generally, we only need the ratio $\dfrac{\nu}{T}$ to estimate the gains and errors.

\subsection{Yield $Y_1$}
In the experiment, we define coincident detection, where both detection sites report a single click, as a successful detection event. A double-click event at one detection site is treated as a no-click event. Additionally, we define cases where only one of the detection sites reports a single click while the other site reports no click as a single detection event. We apply the non-ideal single-photon source model introduced in the previous subsection. For simplicity, we consider a symmetric setting, where the channel loss at Alice's and Bob's sides is the same, i.e., $\eta_1=\eta_4$ and $\eta_2=\eta_3$. Note that $\eta_1$ and $\eta_2$ are not necessarily equal.

According to the security analysis and simulation method in \cite{Ma2018phase}, we need to calculate $Y_1$, the probability of obtaining a successful detection given that Alice and Bob emit a total of one photon. First, we consider the case of an ideal successful detection event, which occurs when both clicks at the two measurement sites are caused by single photons, with one photon originating from the single-photon source and the other from the users. The single photon from the single-photon source can go into either $A_2$ or $B_2$, and the single photon from the users can go into either $A_1$ or $B_1$. Thus, there are four possible cases: $A_1A_2$, $A_1B_2$, $B_1A_2$, $B_1B_2$, each occurring with a probability of $\frac{1}{4}$.

Considering the ideal successful detection event, only the $A_1B_2$ or $B_1A_2$ cases will result in successful detection, so:
\begin{equation}
	\begin{split}
		Y_1^{\text{ideal}} &= \dfrac{1}{4} T\eta_1\eta_3 + \dfrac{1}{4} T\eta_2\eta_4 \\
		&= \dfrac{1}{2} T\eta_1\eta_2,
	\end{split}
\end{equation}
considering that no photon is lost in the channel. In general, the parameter $T$ can be absorbed into the channel transmittance $\eta_2$ or $\eta_3$ because Alice and Bob cannot distinguish whether the light source is not emitting a photon or if the photon is lost in the channel. Thus, we treat $T$ as another form of channel transmittance. Here, we retain $T$ to make the physical meaning clearer.

Next, we consider the case of ideal single detection events, where the two photons from the single-photon source and the users lead to only one click. Similarly, there are four possible cases for the paths of the two photons: $A_1A_2$, $A_1B_2$, $B_1A_2$, and $B_1B_2$. The ideal single detection event (SDE) occurs when one of the two photons is lost in the channel while the other reaches the detection site successfully, or both photons arrive at the same detection site. Thus, we have:
\begin{equation}
	\begin{split}
		\Pr(\text{SDE}) &= \dfrac{1}{4} \eta_1(1-T\eta_3) \qquad \text{(for $A_1B_2$, $B_2$ lost)} \\
		&+ \dfrac{1}{4} (1-\eta_1)T\eta_3 \qquad \text{(for $A_1B_2$, $A_1$ lost)} \\
		&+ \dfrac{1}{4} T\eta_2(1-\eta_4) \qquad \text{(for $A_2B_1$, $B_1$ lost)} \\
		&+ \dfrac{1}{4} (1-T\eta_2)\eta_4 \qquad \text{(for $A_2B_1$, $A_2$ lost)} \\
		&+ \dfrac{1}{4} \eta_1(1-T\eta_2) \qquad \text{(for $A_1A_2$, $A_2$ lost)} \\
		&+ \dfrac{1}{4} (1-\eta_1)T\eta_2 \qquad \text{(for $A_1A_2$, $A_1$ lost)} \\
		&+ \dfrac{1}{4} \eta_1T\eta_2 \qquad \text{(for $A_1A_2$, no loss)} \\
		&+ \dfrac{1}{4} \eta_4(1-T\eta_3) \qquad \text{(for $B_1B_2$, $B_2$ lost)} \\
		&+ \dfrac{1}{4} (1-\eta_4)T\eta_3 \qquad \text{(for $B_1B_2$, $B_1$ lost)} \\
		&+ \dfrac{1}{4} \eta_4T\eta_3 \qquad \text{(for $B_1B_2$, no loss)}.
	\end{split}
\end{equation}
Therefore,
\begin{equation}
	\Pr(\text{SDE}) = \eta_1(1-T\eta_2) + (1-\eta_1)T\eta_2 + \dfrac{1}{2} \eta_1T\eta_2.
\end{equation}

Now we consider non-ideal detection events, in which one detection is caused by a single-photon state and the other is caused by either a dark count or the leaked coherent state on mode $c$. We ignore the case where both detection events are caused by dark counts and leaked coherent states, as its probability is significantly lower than the other cases. 

Denote the dark count rate of a single detector as $p_d$. Since there are two detectors at each detection site, the probability of reporting a click event due to a dark count is $2p_d$ (omitting higher-order terms). Therefore, the probability of non-ideal detection events caused by dark counts is:
\begin{equation}
	Y_1^{DC} = (2p_d)\Pr(\text{SDE}).
\end{equation}

For non-ideal detection events caused by the leaked coherent state on mode $c$, after passing through the beam splitter, the intensity of the coherent state becomes $\nu/2$ when entering channels $A_2$ and $B_2$. Since the single photon from Alice's or Bob's side is in mode $s$ and does not interfere with the coherent state on mode $c$, the probability of the coherent state causing a click on a single side is $1-e^{-1/2\eta_2\nu}$. Thus, the probability of non-ideal successful detection events caused by the coherent state is:
\begin{equation}
	Y_1^{\text{leak}} = (1-e^{-1/2\eta_2\nu})\Pr(\text{SDE}).
\end{equation}

Finally, the total yield $Y_1$ is the sum of the aforementioned three probabilities:
\begin{equation}
	Y_1 = Y_1^{\text{ideal}} + Y_1^{DC} + Y_1^{\text{leak}}.
\end{equation}
The first term in $Y_1$ corresponds to valid detection events, while the other two terms contribute a 50\% error rate.

\subsection{Gain $Q_{\mu}$ and Error Rate $E_{\mu}$}
The gain $Q_{\mu}$ is defined as the probability of obtaining a successful detection event given that Alice and Bob send a total intensity of $\mu$. The corresponding error rate is denoted as $E_{\mu}$. In practice, both values can be directly obtained from experiments. Here, we show the simulation model for calculating these values. Similar to the previous subsection, we consider ideal and non-ideal detection events separately.

To determine $Q_{\mu}$, we first consider the ideal successful detection event. The intensity of pulses emitted by Alice and Bob is $\mu/2$. The single photon emitted by the single-photon source has a $50\%$ chance of going into channel $A_2$ or $B_2$. Considering one photon reaching each detection site, the probability of an ideal successful detection event is:
\begin{equation}
	\begin{split}
		Q_{\mu}^{\text{ideal}} &= \dfrac{1}{2} \left[1-e^{-\eta_1/2\mu}(1-T\eta_2)\right](1-e^{-\eta_4/2\mu})+\dfrac{1}{2} \left[1-e^{-\eta_4/2\mu}(1-T\eta_3)\right](1-e^{-\eta_1/2\mu}) \\
		&= \left[1-e^{-1/2\eta_1\mu}(1-T\eta_2)\right](1-e^{-1/2\eta_1\mu}).
	\end{split}
\end{equation}

Next, we consider the case of ideal single detection events:
\begin{equation}
	\begin{split}
		\Pr(\text{SDE}, \mu) &= \dfrac{1}{2} e^{-\eta_1/2\mu}(1-T\eta_2)(1-e^{-\eta_4/2\mu}) + \dfrac{1}{2} \left[1-e^{-\eta_1/2\mu}(1-T\eta_2)\right]e^{-\eta_4/2\mu} \\
		&+ \dfrac{1}{2} e^{-\eta_4/2\mu}(1-T\eta_3)(1-e^{-\eta_1/2\mu}) + \dfrac{1}{2} \left[1-e^{-\eta_4/2\mu}(1-T\eta_3)\right]e^{-\eta_1/2\mu} \\
		&= e^{-1/2\eta_1\mu}(1-T\eta_2)(1-e^{-1/2\eta_1\mu}) + \left[1-e^{-1/2\eta_1\mu}(1-T\eta_2)\right]e^{-1/2\eta_1\mu}.
	\end{split}
\end{equation}

Then, we consider the non-ideal successful detection events in which one of the detections is an ideal single detection, while the other is caused by either a dark count or the leaked coherent state on mode $c$. Similar to $Y_1$, the probabilities of these two non-ideal cases are given by:
\begin{equation}
	\begin{split}
		Q_{\mu}^{\text{dark}} &= (2p_d) \Pr(\text{SDE}, \mu), \\
		Q_{\mu}^{\text{leak}} &= (1-e^{-1/2\eta_2\nu})\Pr(\text{SDE}, \mu).
	\end{split}
\end{equation}

Therefore, the total $Q_{\mu}$ is:
\begin{equation}
	Q_{\mu} = Q_{\mu}^{\text{ideal}} + Q_{\mu}^{\text{dark}} + Q_{\mu}^{\text{leak}}.
\end{equation}

The error rate $E_{\mu}$ consists of three parts: errors from the ideal case (similar to Eqs.~\eqref{eq:error0} and \eqref{eq:errorpi}), errors caused by dark counts, and errors caused by the leaked coherent state from the single-photon source. The latter two parts introduce totally random raw key bits, leading to an error rate of $0.5$. The first part depends the size of the random phase slices, $\frac{2\pi}{D}$, and the average error rate of this part can be estimated as:
\begin{equation}
	\begin{split}
		e_{\mu}^{\text{ideal}} &\approx \dfrac{\eta_1\eta_4(2-\eta_2-\eta_3)\abs{\alpha\beta}^2}{\abs{\sqrt{\eta_1\eta_3}\alpha + \sqrt{\eta_2\eta_4}\beta}^2 + 2\eta_1\eta_4(2-\eta_2-\eta_3)\abs{\alpha\beta}^2} \\
%		&\lesssim \dfrac{\eta_4(2-\eta_2-\eta_3)\mu}{\eta_3\abs{e^{-i\theta}+e^{-i(\theta+\frac{\pi}{D})}}^2 + 2\eta_4(2-\eta_2-\eta_3)\mu}\\
		&\lesssim \dfrac{\eta_4(2-\eta_2-\eta_3)\mu}{4\eta_3 \cos(\dfrac{\pi}{2D})^2 + 2\eta_4(2-\eta_2-\eta_3)\mu},
	\end{split}
\end{equation}
where the second approximate equality holds when $D$ is large. In our experiment, we set $D=16$, which provides a good approximation.

Thus, the error rate $E_{\mu}$ is:
\begin{equation}
	E_{\mu} = \dfrac{e_{\mu}^{\text{ideal}} Q_{\mu}^{\text{ideal}} + 0.5 Q_{\mu}^{\text{dark}} + 0.5 Q_{\mu}^{\text{leak}}}{Q_{\mu}}.
\end{equation}

The simulation method primarily follows the phase-matching scheme from \cite{Ma2018phase}. The phase error rate can be upper bounded by $Y_1$ and $Q_{\mu}$ as follows:
\begin{equation}
	\begin{split}
		q_1 &= Y_1 \dfrac{\mu e^{-\mu}}{Q_{\mu}}, \\
		E_{\mu}^{\text{ph}} &\leq 1 - q_1.
	\end{split}
\end{equation}
The key rate is then calculated as:
\begin{equation}
	r = \dfrac{2Q_{\mu}}{D} \left( 1 - h(E_{\mu}^{\text{ph}}) - f h(E_{\mu}) \right),
\end{equation}
where $h(x) = -x\log(x) - (1-x)\log(1-x)$ is the binary entropy function.

\subsection{Simulation Results}
Using the model described above, we simulated the key generation rate for our five-node setup in the asymptotic limit. To maximize the key generation rate, we optimized the light intensities sent by Alice and Bob, as well as the transmission distances between the nodes. It is important to note that the simulation only considers error rates originating from the protocol itself and errors introduced by the leaked coherent light in the single-photon source.

\begin{figure}[hbtp!]
	\centering
	\includegraphics[width=0.5\linewidth]{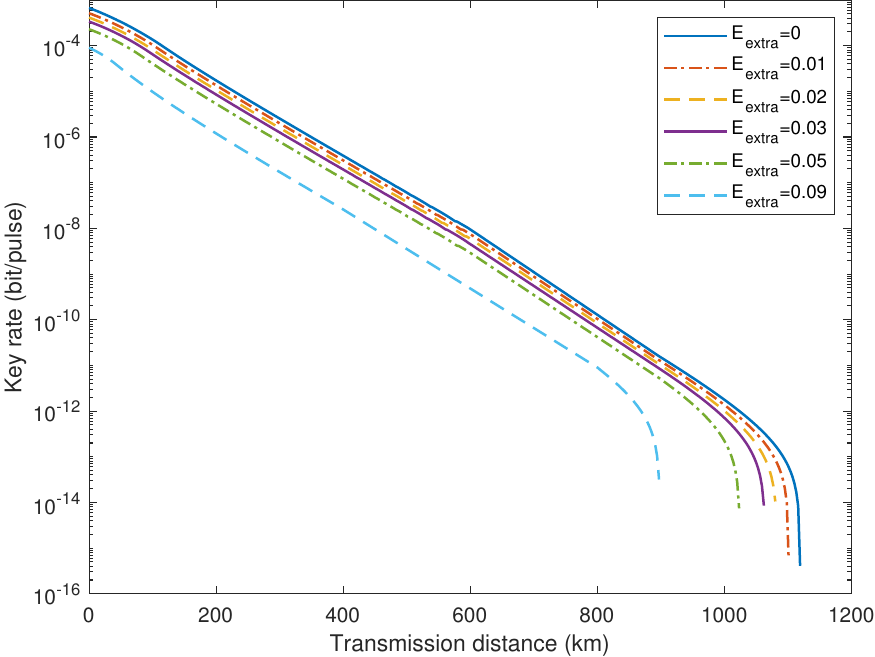}
	\caption{Simulation results. We consider a single-photon source with $g^{(2)}(0) = 0.002$ and an efficiency of 0.3, and calculate the key generation rate for additional error rates $E_{\text{extra}}$ of 0\%, 1\%, 2\%, 3\%, 5\%, and 9\%. The light intensities and distances between the nodes are optimized for different transmission distances.}
	\label{fig:simresult}
\end{figure}

In our actual experiments, we observed that interference quality also affects the overall error rate. The quality of interference is impacted by various factors, including light intensity, modulation accuracy, synchronization precision, and channel disturbances. To account for these effects in our simulation, we introduced an additional error rate, $E_{\text{extra}}$. The simulation results are shown in Fig.~\ref{fig:simresult}.

From the simulation results, it is evident that, with a low extra error rate, our setup can successfully distribute keys over distances beyond 1000 km. The current experimental system exhibits an additional error rate of approximately 9\%. Therefore, improving the phase stability will be crucial for enabling longer-distance transmission in the future. This represents an interesting direction for further research and development.

\begin{figure}[hbtp!]
	\centering
	\includegraphics[width=0.5\linewidth]{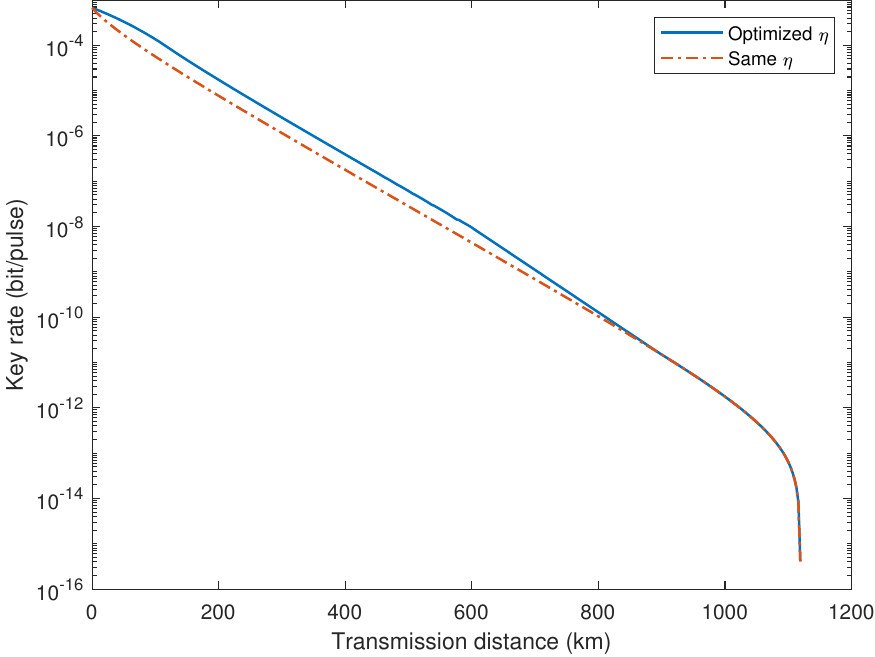}
	\caption{Comparison between results with and without optimization of the losses between nodes. We consider the case where $E_{\text{extra}} = 0$. The results show that optimizing the loss distribution between nodes can enhance system performance. }
	\label{fig:optcompare}
\end{figure}

In Fig.~\ref{fig:optcompare}, we compare simulation results with and without optimization of the losses between nodes, assuming no additional errors in both cases. The results indicate that optimizing the distances between nodes can indeed improve system performance. Determining how to implement such adjustments and optimize the specific distances and light intensities remains an important direction for future research.

\begin{figure}[hbtp!]
	\centering
	\includegraphics[width=0.5\linewidth]{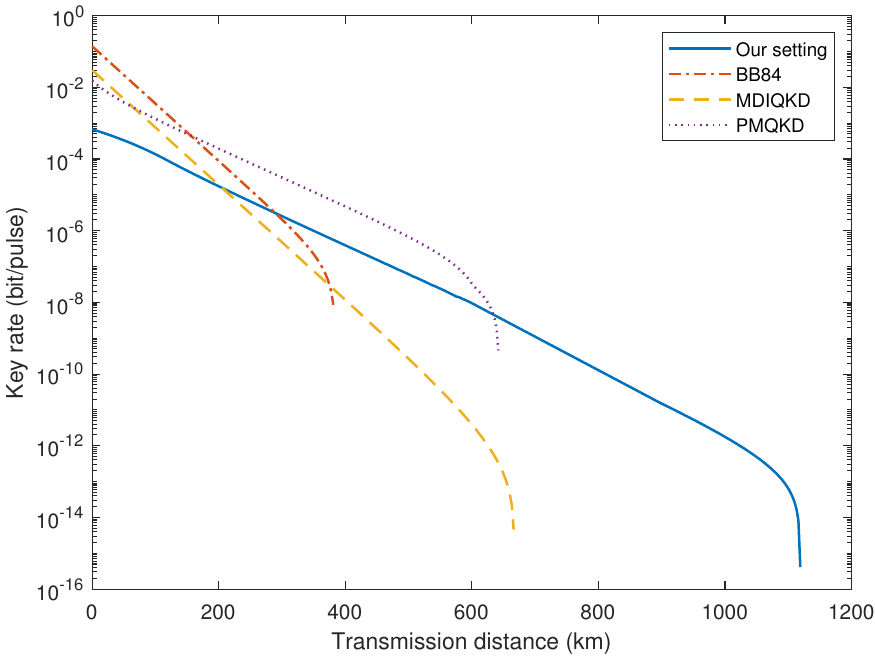}
	\caption{Comparison between different protocols. All protocols are simulated under ideal conditions without considering finite key lengths, with the sole limitation being the detector dark count rate, \(p_d = 2.78 \times 10^{-8}\), consistent with the value used in our experiments. The results demonstrate that applying QKD protocols to our five-node relay structure can extend the maximum communication distance to nearly twice that of MDI protocols.}
	\label{fig:protocolcompare}
\end{figure}

Finally, we compared the performance of QKD protocols implemented in our system with other established QKD protocols, such as BB84, measurement-device-independent (MDI) protocol, and phase-matching protocol in Fig.~\ref{fig:protocolcompare}. The results indicate that when applied to our five-node relay structure, the scaling of the QKD protocol matches that of twin-field-like protocols, at \(O(\sqrt{\eta})\). Moreover, due to the improved signal-to-noise ratio in individual links, our scheme extends the maximum communication distance to nearly twice that of MDI protocols.

\section{Relationship between Visibility and Errors}\label{sc:tpi}
As mentioned in the main text, imperfect interference between single photons and coherent light, quantified by interference visibility, introduces additional errors. In this section, we provide a theoretical analysis of this relationship and compare it with the experimental results.

\subsection{Errors Introduced by Imperfect Interference}\label{app:error}
In practice, the photon emitted by the single-photon source is not identical to the laser pulse, leading to imperfect interference and potentially increasing the error rate. This difference can be a significant source of error. Here, we calculate the error rate due to the imperfection of the optical interference. 

We assume that the single-photon mode that can interfere with the laser is $\ket{1}$, while the mode that cannot interfere is $\ket{1'}$. Their emission probabilities are $p_I$ and $p_N$, respectively, and their creation operators are denoted by $a^\dagger$ and $A^\dagger$. Thus, the photon emitted by the single-photon source can be expressed as:
\begin{equation}
	\sqrt{p_I}\ket{1} + \sqrt{p_N}\ket{1'} + \sqrt{1-p_I-p_N}\ket{0} = (\sqrt{p_I} a^\dagger + \sqrt{p_N} A^\dagger + \sqrt{1-p_I-p_N}) \ket{0}.
\end{equation}

For simplicity, we assume that the phase difference of the coherent states of Alice and Bob is 0 in the following derivation. When the phase difference is $\pi$, the results are the same, with only the interpretation of the click event changing according to the protocol. The emitted state can be written as:
\begin{equation}
	D_{A1}(\alpha) D_{B1}(\alpha) \left(\sqrt{p_I} a^\dagger_s + \sqrt{p_N} A^\dagger_s + \sqrt{1 - p_I - p_N} I \right) \ket{0}
\end{equation}
where the displacement operator is given by $D_{A1}(\alpha) = e^{-\abs{\alpha}^2/2}e^{\alpha a_1^\dagger}$. 

After passing through the beam splitter in front of the single-photon source, the state changes to:
\begin{equation}
	D_{A1}(\alpha) D_{B1}(\alpha) \left( \sqrt{\frac{p_I}{2}} a^\dagger_{A2} + \sqrt{\frac{p_I}{2}} a^\dagger_{B2} + \sqrt{\frac{p_N}{2}} A^\dagger_{A2} + \sqrt{\frac{p_N}{2}} A^\dagger_{B2} + \sqrt{1 - p_I - p_N} I \right) \ket{0}.
\end{equation}

Assuming a channel transmittance of $\eta$, the state after passing through the channel becomes:
\begin{equation}
	D_{A1}(\sqrt{\eta}\alpha) D_{B1}(\sqrt{\eta}\alpha) \left( \sqrt{\frac{\eta p_I}{2}} a^\dagger_{A2} + \sqrt{\frac{\eta p_I}{2}} a^\dagger_{B2} + \sqrt{\frac{\eta p_N}{2}} A^\dagger_{A2} + \sqrt{\frac{\eta p_N}{2}} A^\dagger_{B2} + \sqrt{1 - \eta p_I - \eta p_N} I \right) \ket{0}.
\end{equation}

After interference on the beam splitters in front of the single-photon detectors, the state changes to:
\begin{equation}
	\begin{aligned}
		& D_{L1} \left( \sqrt{\frac{\eta}{2}}\alpha \right) D_{R1} \left( -\sqrt{\frac{\eta}{2}}\alpha \right) D_{L2} \left( \sqrt{\frac{\eta}{2}}\alpha \right) D_{R2} \left( -\sqrt{\frac{\eta}{2}}\alpha \right) \sqrt{1 - \eta p_I - \eta p_N} I \\
		& + \sqrt{\frac{\eta p_I}{2}} D_{L1} \left( \sqrt{\frac{\eta}{2}}\alpha \right) D_{R1} \left( -\sqrt{\frac{\eta}{2}}\alpha \right) D_{L2} \left( \sqrt{\frac{\eta}{2}}\alpha \right) D_{R2} \left( -\sqrt{\frac{\eta}{2}}\alpha \right) \left( \sqrt{\frac{1}{2}}a^\dagger_{L1} + \sqrt{\frac{1}{2}}a^\dagger_{R1} \right) \\
		& + \sqrt{\frac{\eta p_I}{2}} D_{L1} \left( \sqrt{\frac{\eta}{2}}\alpha \right) D_{R1} \left( -\sqrt{\frac{\eta}{2}}\alpha \right) D_{L2} \left( \sqrt{\frac{\eta}{2}}\alpha \right) D_{R2} \left( -\sqrt{\frac{\eta}{2}}\alpha \right) \left( \sqrt{\frac{1}{2}}a^\dagger_{L2} + \sqrt{\frac{1}{2}}a^\dagger_{R2} \right) \\
		& + \sqrt{\frac{\eta p_N}{2}} D_{L1} \left( \sqrt{\frac{\eta}{2}}\alpha \right) D_{R1} \left( -\sqrt{\frac{\eta}{2}}\alpha \right) D_{L2} \left( \sqrt{\frac{\eta}{2}}\alpha \right) D_{R2} \left( -\sqrt{\frac{\eta}{2}}\alpha \right) \left( \sqrt{\frac{1}{2}}A^\dagger_{L1} + \sqrt{\frac{1}{2}}A^\dagger_{R1} \right) \\
		& + \sqrt{\frac{\eta p_N}{2}} D_{L1} \left( \sqrt{\frac{\eta}{2}}\alpha \right) D_{R1} \left( -\sqrt{\frac{\eta}{2}}\alpha \right) D_{L2} \left( \sqrt{\frac{\eta}{2}}\alpha \right) D_{R2} \left( -\sqrt{\frac{\eta}{2}}\alpha \right) \left( \sqrt{\frac{1}{2}}A^\dagger_{L2} + \sqrt{\frac{1}{2}}A^\dagger_{R2} \right) \ket{0}
	\end{aligned}
\end{equation}
Since $\eta p \ll 1$ and $\eta \alpha^2 \ll 1$, we ignore higher-order terms, and the state becomes:
\begin{equation}
	\begin{aligned}
		& \sqrt{1 - \eta p_I - \eta p_N} \left( I + \sqrt{\frac{\eta}{2}}\alpha a^\dagger_{L1} \right) \left( I - \sqrt{\frac{\eta}{2}}\alpha a^\dagger_{R1} \right) \left( I + \sqrt{\frac{\eta}{2}}\alpha a^\dagger_{L2} \right) \left( I - \sqrt{\frac{\eta}{2}}\alpha a^\dagger_{R2} \right) \\
		& + \sqrt{\frac{\eta p_I}{2}} \left( I + \sqrt{\frac{\eta}{2}}\alpha a^\dagger_{L1} \right) \left( I - \sqrt{\frac{\eta}{2}}\alpha a^\dagger_{R1} \right) \left( I + \sqrt{\frac{\eta}{2}}\alpha a^\dagger_{L2} \right) \left( I - \sqrt{\frac{\eta}{2}}\alpha a^\dagger_{R2} \right) \left( \sqrt{\frac{1}{2}}a^\dagger_{L1} + \sqrt{\frac{1}{2}}a^\dagger_{R1} \right) \\
		& + \sqrt{\frac{\eta p_N}{2}} \left( I + \sqrt{\frac{\eta}{2}}\alpha a^\dagger_{L1} \right) \left( I - \sqrt{\frac{\eta}{2}}\alpha a^\dagger_{R1} \right) \left( I + \sqrt{\frac{\eta}{2}}\alpha a^\dagger_{L2} \right) \left( I - \sqrt{\frac{\eta}{2}}\alpha a^\dagger_{R2} \right) \left( \sqrt{\frac{1}{2}}A^\dagger_{L1} + \sqrt{\frac{1}{2}}A^\dagger_{R1} \right) \\
		& + \sqrt{\frac{\eta p_N}{2}} \left( I + \sqrt{\frac{\eta}{2}}\alpha a^\dagger_{L1} \right) \left( I - \sqrt{\frac{\eta}{2}}\alpha a^\dagger_{R1} \right) \left( I + \sqrt{\frac{\eta}{2}}\alpha a^\dagger_{L2} \right) \left( I - \sqrt{\frac{\eta}{2}}\alpha a^\dagger_{R2} \right) \left( \sqrt{\frac{1}{2}}A^\dagger_{L2} + \sqrt{\frac{1}{2}}A^\dagger_{R2} \right) \ket{0}
	\end{aligned}
\end{equation}
Retaining only events where there is one click on each side, the resulting rate of correct events is:
\begin{equation}
	(1 - \eta p_I - \eta p_N)\eta^2 \alpha^4 / 2 + \eta^2 p_I \alpha^2 + \eta^2 p_N \alpha^2 / 2
\end{equation}
The rate of error events is:
\begin{equation}
	(1 - \eta p_I - \eta p_N)\eta^2 \alpha^4 / 2 + \eta^2 p_N \alpha^2 / 2
\end{equation}
The error rate is:
\begin{equation}\label{eq:e}
	e = \frac{(1 - \eta p_I - \eta p_N)\alpha^2 + p_N}{2(1 - \eta p_I - \eta p_N)\alpha^2 + 2p_I + 2p_N}
\end{equation}

To determine $p_N$, we use the experiment results of the laser and single-photon source. The interference visibility is assumed to be $V$, and the state before the beam splitter is:
\begin{equation}
	\left( \sqrt{p_0} I + \sqrt{p_I} a^\dagger_0 + \sqrt{p_N} A^\dagger_0 \right) D_1(\alpha) \ket{0}
\end{equation}
After interference, it becomes:
\begin{equation}
	\begin{aligned}
		&\sqrt{p_0} D_2 \left( \sqrt{\frac{1}{2}}\alpha \right) D_3 \left( -\sqrt{\frac{1}{2}}\alpha \right)\\
		& + \sqrt{\frac{p_I}{2}} D_2 \left( \sqrt{\frac{1}{2}}\alpha \right) D_3 \left( -\sqrt{\frac{1}{2}}\alpha \right)a^\dagger_2
		+ \sqrt{\frac{p_I}{2}} D_2 \left( \sqrt{\frac{1}{2}}\alpha \right) D_3 \left( -\sqrt{\frac{1}{2}}\alpha \right)a^\dagger_3\\
		& + \sqrt{\frac{p_I}{2}} D_2 \left( \sqrt{\frac{1}{2}}\alpha \right) D_3 \left( -\sqrt{\frac{1}{2}}\alpha \right)A^\dagger_2
		+ \sqrt{\frac{p_I}{2}} D_2 \left( \sqrt{\frac{1}{2}}\alpha \right) D_3 \left( -\sqrt{\frac{1}{2}}\alpha \right)A^\dagger_3\ket{0}\\
	\end{aligned}
	\nonumber
\end{equation}
Assuming that both the coherent state and the single photon light intensity are weak enough so we can expand expression to
\begin{equation}
	\left(-\frac{\sqrt{p_0}}{2}\alpha^2 a^\dagger_2 a^\dagger_3 -\frac{\sqrt{p_N}}{2}\alpha A^\dagger_2 a^\dagger_3 + \frac{\sqrt{p_N}}{2}\alpha a^\dagger_2 A^\dagger_3\right)\ket{0}.
\end{equation}

The interference visibility can be expressed as:
\begin{equation}
	V = 1 - \frac{\frac{p_0}{4}\alpha^4 + \frac{p_N}{2} \alpha^2}{\frac{p_0}{4}\alpha^4 + \frac{1 - p_0}{2}\alpha^2}
\end{equation}
From this, we have:
\begin{equation}\label{eq:pN}
	p_N = \frac{(1 - V)\left(\frac{p_0}{4}\alpha^4 + \frac{1 - p_0}{2}\alpha^2\right) - \frac{p_0}{4}\alpha^4}{\frac{\alpha^2}{2}}
\end{equation}
Combining Eqs.~\eqref{eq:pN} and \eqref{eq:e}, we derive the relationship between the error rate and interference visibility. It is important to note that in our experimental setup, the single-photon source passes through a beam splitter, and thus the values of $p_I$ and $p_N$ for calculating the interference visibility are half of those used for calculating the error rate. In subsequent equations, $p_0$, $p_I$, and $p_N$ represent photon yields observed in the interference. The final result is as follows:
\begin{equation}\label{eq:errv}
	e = \frac{2[1 - 2\eta(1 - p_0)]\alpha^2 + 2(1 - p_0)(1 - V) - p_0\alpha^2 V}{4[1 - 2\eta(1 - p_0)]\alpha^2 + 4(1 - p_0)}.
\end{equation}

\subsection{Experimental Results}
\begin{figure}[hbpt!]
	\includegraphics[width=0.5\linewidth]{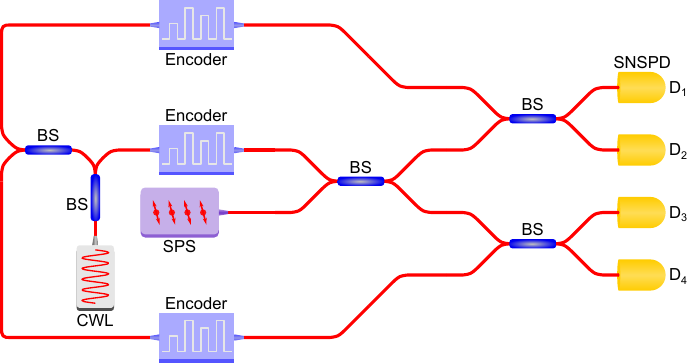}
	\caption{Interference visibility test setup. CWL: continuous-wave laser; SPS: single-photon source; BS: beam splitter; SNSPD: superconducting nanowire single-photon detector.}
	\label{fig:testsetting}
\end{figure}

To minimize the influence of factors other than interference visibility on the error rate, we conduct an experiment to test the system's error rate at a repetition rate of $\SI{76.13}{\MHz}$. As illustrated in Fig.~\ref{fig:testsetting}, we use a single continuous-wave laser to generate three light beams instead of using three separate lasers (i.e., a master continuous-wave laser and two slave continuous-wave lasers). The setup connects two user nodes or the quantum relay node directly with the measurement nodes.

In the signal pulse region, signal states with $D=16$ discrete random phases are modulated. The system runs for over half an hour, collecting 60 sets of data, each corresponding to 30 seconds of measurement. We process the collected data using different detection time windows to evaluate the error rates.

\begin{figure}[hbpt!]
	\includegraphics[width=\linewidth]{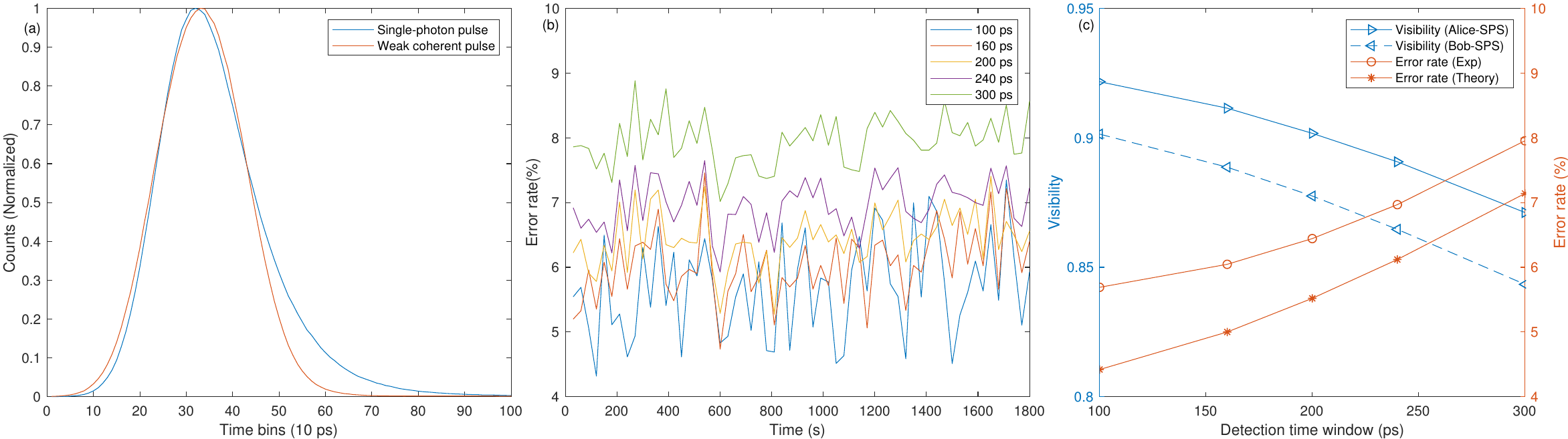}
	\caption{(a) Pulse shapes of weak coherent pulses and single photons. (b) Error rates obtained from 60 sets of data, correspond to the detection time windows of $\SI{100}{\ps}$, $\SI{160}{\ps}$, $\SI{200}{\ps}$, $\SI{240}{\ps}$, and $\SI{300}{\ps}$, respectively. (c) Total interference visibility between Alice or Bob's coherent pulses and single photons, and the corresponding total error rate for different detection time windows.}
	\label{fig:erver}
\end{figure}

As shown in Fig.~\ref{fig:erver} (a), the pulse shapes of coherent pulses and single photons obtained by statistical counting of SNSPD are different due to different generation mechanisms. Thus, reducing the detection time window improves the interference quality between coherent pulses and single photons, resulting in a lower error rate. However, a smaller detection time window also leads to a greater variation in the error rate due to a shorter raw key length, as depicted in Fig.~\ref{fig:erver} (b). To quantify the interference quality between the coherent state and the single-photon source, we use the same data to estimate the interference visibility.

%\begin{figure}[hbpt!]
%	\includegraphics[width=\linewidth]{figerver4.pdf}
%	\caption{\red{(a) Pulse shapes of weak coherent pulses and single photons. (b) Error rates obtained from 60 sets of data, correspond to half of the detection time windows of $\SI{50}{\ps}$, $\SI{80}{\ps}$, $\SI{100}{\ps}$, $\SI{120}{\ps}$, and $\SI{150}{\ps}$, respectively. (c) Total interference visibility between Alice or Bob's coherent pulses and single photons, and the corresponding total error rate for different detection time windows.}}
%	\label{fig:erver}
%\end{figure}
The interference visibility between the coherent pulses from Alice's and Bob's lasers and the single-photon source, denoted as $V_A$ and $V_B$, is calculated as:
\begin{equation}
	\begin{split}
		V_A &= 1 - \frac{P_{C(1, 2)}}{P'_{C(1, 2)}}, \\
		V_B &= 1 - \frac{P_{C(3, 4)}}{P'_{C(3, 4)}},
	\end{split}
\end{equation}
where $P_{C(1, 2)}$ ($P_{C(3, 4)}$) represents the probability of simultaneous clicks at D\textsubscript{1} (D\textsubscript{3}) and D\textsubscript{2} (D\textsubscript{4}) for the same mode, while $P'_{C(1, 2)}$ ($P'_{C(3, 4)}$) represents the corresponding probability for different modes. $P_{C(1, 2)}$ and $P_{C(3, 4)}$ are obtained directly from the experimental data, while $P'_{C(1, 2)}$ and $P'_{C(3, 4)}$ are derived from the following formulas:
\begin{equation}
	\begin{split}
		P'_{C(1, 2)} &= P^\text{coh}_1 P^\text{coh}_2 + P^\text{sp}_1 P^\text{sp}_2 g^2(0) + P^\text{coh}_1 P^\text{sp}_2 + P^\text{coh}_2 P^\text{sp}_1, \\
		P'_{C(3, 4)} &= P^\text{coh}_3 P^\text{coh}_4 + P^\text{sp}_3 P^\text{sp}_4 g^2(0) + P^\text{coh}_3 P^\text{sp}_4 + P^\text{coh}_4 P^\text{sp}_3,
	\end{split}
\end{equation}
where $P^\text{coh}_i$ ($P^\text{sp}_i$) represents the detection probability at detector D\textsubscript{i} when only the coherent state (single-photon) is sent.

Using these probabilities, we estimate $V_A$ and $V_B$ for different detection time windows, as shown in Fig.~\ref{fig:erver} (c). The reason for the difference in visibility of $V_A$ and $V_B$ is that the shapes of the coherent pulses generated by Alice and Bob are not exactly the same, as well as the splitting ratios of the beam splitters used for interference. Substituting the average values of $V_A$ and $V_B$ into Eq.~\eqref{eq:errv} provides the theoretical error rates for different detection time windows. We find that the experimental error rate is approximately $1\%$ higher than the theoretical prediction, which we attribute to phase modulation errors.

\section{Phase Reference}\label{sc:OPLL}
In this section, we describe how we establish the phase reference between the single-photon source and the laser. Specifically, we explain our phase-locking method and phase estimation method, as a supplement to the Methods section in the main text.

\subsection{Phase Locking}
To satisfy the requirements of the phase-matching quantum key distribution protocol, a stable phase reference must be maintained between the lasers of the two users. We achieve this stability using heterodyne optical phase-locked loop (OPLL) technology \cite{minder2019experimental}, ensuring that the two slave continuous-wave lasers (SCWLs) have the same frequency and a relatively stable initial phase difference.

Alice's and Bob's SCWLs are fiber lasers whose wavelengths can be adjusted by either changing the temperature of the laser substrate or by applying tensile strain on the laser cavity using a piezoelectric actuator. The continuous-wave light from the local laser is frequency-shifted by an acousto-optic modulator (AOM) at approximately $\SI{40}{\MHz}$. A portion of the light interferes with the light from the quantum relay site and is detected by a balanced detector (BD).

\begin{figure}[hbpt!]
	\includegraphics[width=0.5\linewidth]{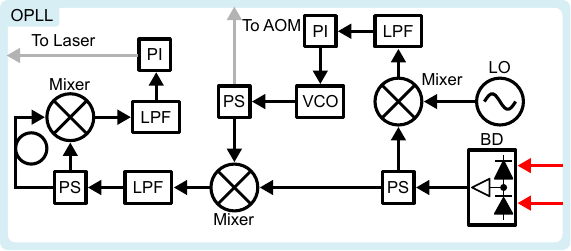}
	\caption{\label{fig:opll} Heterodyne optical phase-locked loop (OPLL) technology. BD: balanced detector; PS: power splitter; LO: local oscillator; LP: loop filter; LFP: low-pass filter; PI: proportional-integral controller; VCO: voltage-controlled oscillator.}
\end{figure}

As shown in Fig.~\ref{fig:opll}, the electrical signal output of the BD at frequency $f_\text{beat}$ is split into two parts by a power splitter. One part of the signal is used for low-speed, large-range feedback, which controls the frequency of the laser via the piezoelectric actuator. The other part is used for high-speed, small-range feedback, with the feedback signal controlling the frequency of the AOM shift via a voltage-controlled oscillator (VCO).

The electrical signal used for slow feedback is mixed with a portion of the VCO output at frequency $f_\text{VCO}$. The resulting mixed signal passes through a low-pass filter (LPF) to generate a signal with a frequency difference of $f_\text{beat} - f_\text{VCO}$. This signal is divided into two paths, and after a delay of $\Delta t = \SI{25}{\ns}$, it is mixed with itself and then filtered by the LPF. The resulting error signal is approximately a direct current (DC) value proportional to $\cos (2\pi(f_\text{beat} - f_\text{VCO})\Delta t)$. When $f_\text{beat} - f_\text{VCO} = 10 \times (2n + 1)\SI{}{\MHz}$, where $n \geq 0$ is an integer, the error signal becomes zero.

\begin{figure}[hbpt!]
	\includegraphics[width=0.6\linewidth]{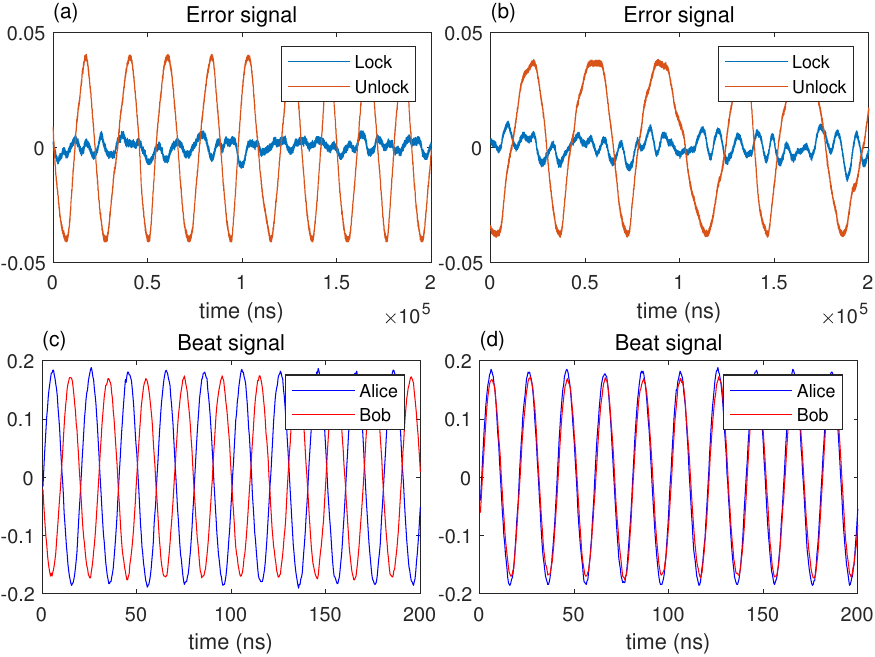}
	\caption{Error signal and beat signal. (a) Error signals when Alice's laser is locked and unlocked; (b) Error signals when Bob's laser is locked and unlocked; (c) Beat signals when Alice and Bob's lasers are unlocked; (d) Beat signals when Alice and Bob's lasers are locked.}
	\label{fig:opllr}
\end{figure}

Under initial conditions, the output frequency of the VCO is approximately $\SI{40}{\MHz}$. Laser frequency control is achieved by temperature tuning such that $f_\text{beat}$ falls between $\SI{40}{\MHz}$ and $\SI{60}{\MHz}$. The proportional-integral (PI) parameter is then adjusted to lock the system, reducing the error signal, resulting in $f_\text{beat} = \SI{50}{\MHz} \pm \Delta f$, where $\Delta f < \SI{0.1}{\MHz}$, as illustrated in Fig.~\ref{fig:opllr}(c).

In this state, the other signal used for fast feedback is mixed with a $\SI{50}{\MHz}$ signal generated by the local oscillator (LO). The resulting error signal after low-pass filtering is $\Delta f$. Adjusting the PI parameters initiates the locking process, keeping the error signal close to zero, as shown in Fig.~\ref{fig:opllr}(a) and (b). Consequently, the frequency of the beat signal is nearly equal to $\SI{50}{\MHz}$, as depicted in Fig.~\ref{fig:opllr}(d). Despite the phase-locking efforts, residual phase noise remains, which can cause additional phase modulation errors.

\subsection{Phase Estimation}
\begin{figure}[hbpt!]
	\includegraphics[width=0.6\linewidth]{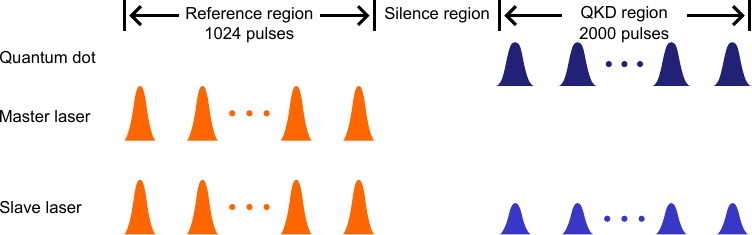}
	\caption{Pulse train design. The QKD system generates 30,452 pulses per $\SI{100}{\us}$, organized into 10 repeating segments excluding the pulses used for time calibration. Each segment consists of a reference pulse region, a silence region, and a QKD region. Each reference pulse region contains 1,024 pulses, and each QKD region contains 2,000 pulses.}
	\label{fig:pulse}
\end{figure}

As described in the method, Alice, Bob, and the quantum relay node each prepare phase reference pulses with initial phases $\theta_A$, $\theta_B$, and $\theta_R$. Every $\SI{100}{\us}$, they generate 10,240 phase reference pulses and 20,000 signal pulses, as shown in Fig.~\ref{fig:pulse}. A portion of the phase reference pulses prepared by Alice and Bob are modulated to have a phase of either $\pi/2$ or $-\pi/2$. This modulation allows us to calculate $\cos(\theta_A + \theta_1 - \theta_R - \theta_2)$, $\sin(\theta_A + \theta_1 - \theta_R - \theta_2)$, $\cos(\theta_B + \theta_4 - \theta_R - \theta_3)$, and $\sin(\theta_B + \theta_4 - \theta_R - \theta_3)$ using the statistical counts $n_{i}^{\phi}$ from four single-photon detectors, where $i=1,\cdots, 4$ and $\phi=0, \pi/2, -\pi/2$ correspond to the phase reference pulses.

\begin{figure}[hbpt!]
	\centering
	\includegraphics[width=0.4\linewidth]{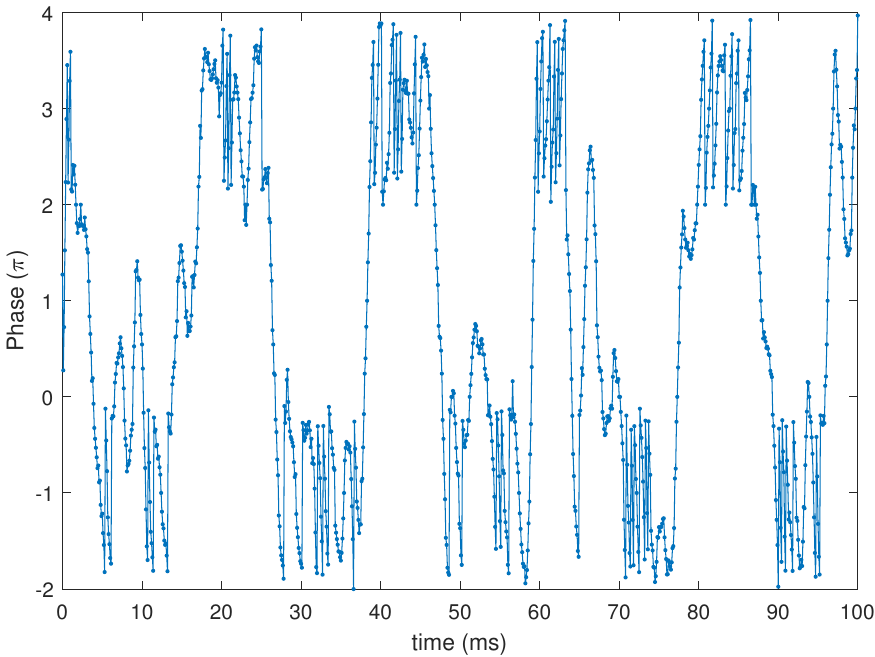}
	\caption{Phase drift over $\SI{100}{\ms}$. The phase drift range is expanded from $\left[0, 2\pi \right)$ to $\left[-2\pi, 4\pi \right)$ to prevent phase jumps.}
	\label{fig:pd}
\end{figure}

With these calculations, we determine $\cos(\Delta \theta)$ and $\sin(\Delta \theta)$, where $\Delta \theta = \theta_A + \theta_1 - \theta_2 - \theta_B - \theta_4 + \theta_3$. Figure \ref{fig:pd} shows the phase drift $\Delta \theta$ over $\SI{100}{\ms}$ caused by a total of $\SI{300}{\km}$ of fiber. To avoid phase jumps, we expand the range of $\Delta \theta$ from $\left[0, 2\pi \right)$ to $\left[-2\pi, 4\pi \right)$.

The estimated phase is used directly for phase compensation in this work. For future improvements, one can employ better estimation methods, such as fitting the estimated phase, especially in regions where rapid phase drift occurs.

\section{Detailed Experimental Data}\label{sc:Exprdata}
In this section, we provide detailed experimental data from our demonstration, summarized in Table \ref{tab:expdata}.

\begin{table}[hbtp!]
	\centering
	\caption{Pulse intensity, channel conditions, and detection results. $N$: Total rounds in which Alice and Bob simultaneously send signal pulses. $N_{00}$: Number of rounds in which Alice and Bob simultaneously send vacuum states. $N_{\nu\nu}$: Number of rounds in which Alice and Bob simultaneously send decoy states. $N_{\mu\mu}$: Number of rounds in which Alice and Bob simultaneously send signal states. $M_{00}$: Effective clicks for rounds with simultaneous vacuum states. $M_{\nu\nu}$: Effective clicks for rounds with simultaneous decoy states. $M_{\mu\mu}$: Effective clicks for rounds with simultaneous signal states.}
	\label{tab:expdata}
	\begin{tabularx}{0.6\linewidth}{c|ccc}
		\hline 
		\textbf{Distance (km)}   & \textbf{100}  & \textbf{200} & \textbf{300} \\  
		\hline 
		Intensity of single photon $\gamma$ & 0.05338 & 0.04954 & 0.05386 \\
		Intensity of signal state $\mu$ & 0.00199 & 0.00199 & 0.00213 \\
		Intensity of decoy state $\nu$ & 0.00080 & 0.00098 & 0.00103 \\
		Total detection efficiency $\eta_d$ & 0.52 & 0.52 & 0.48 \\
		Total channel loss (dB) & 19.34 & 35.58 & 52.33 \\
		$N$ & $9.6\times10^{12}$ & $2.76\times10^{13}$ & $1.56\times10^{14}$  \\  
		$N_{00}$ & $7.776\times 10^{10}$ & $2.829\times 10^{11}$ & $6.786\times10^{12}$ \\
		$N_{\nu\nu}$ & $2.4576\times 10^{11}$ & $2.24112\times 10^{12}$ & $4.24866\times10^{13}$ \\
		$N_{\mu\mu}$ & $5.41488\times 10^{12}$ & $1.06025\times 10^{13}$ & $1.12632\times10^{13}$ \\
		$M_{00}$ & 4588 & 3099 & 12062 \\
		$M_{\nu\nu}$ & 305157 & 518537 & 1424855 \\
		$M_{\mu\mu}$ & 16746919 & 5012015 & 779895 \\
		Raw key length & 2092659 & 627215 & 97169 \\
		Quantum bit error rate & 9.70\% & 9.57\% & 10.44\% \\
		\hline 
	\end{tabularx}
\end{table}

%\bibliographystyle{apsrev}

%\bibliography{bibSPS2}

\begin{thebibliography}{42}
\expandafter\ifx\csname natexlab\endcsname\relax\def\natexlab#1{#1}\fi
\expandafter\ifx\csname bibnamefont\endcsname\relax
  \def\bibnamefont#1{#1}\fi
\expandafter\ifx\csname bibfnamefont\endcsname\relax
  \def\bibfnamefont#1{#1}\fi
\expandafter\ifx\csname citenamefont\endcsname\relax
  \def\citenamefont#1{#1}\fi
\expandafter\ifx\csname url\endcsname\relax
  \def\url#1{\texttt{#1}}\fi
\expandafter\ifx\csname urlprefix\endcsname\relax\def\urlprefix{URL }\fi
\providecommand{\bibinfo}[2]{#2}
\providecommand{\eprint}[2][]{\url{#2}}

\bibitem[{\citenamefont{Shor}(1994)}]{shor1994algorithms}
\bibinfo{author}{\bibfnamefont{P.~W.} \bibnamefont{Shor}}, in
  \emph{\bibinfo{booktitle}{Proceedings 35th Annual Symposium on Foundations of
  Computer Science}} (\bibinfo{year}{1994}), pp. \bibinfo{pages}{124--134}.

\bibitem[{\citenamefont{Grover}(1996)}]{grover1996a}
\bibinfo{author}{\bibfnamefont{L.~K.} \bibnamefont{Grover}}, in
  \emph{\bibinfo{booktitle}{Proceedings of the twenty-eighth annual ACM
  symposium on Theory of computing}} (\bibinfo{year}{1996}), pp.
  \bibinfo{pages}{212--219}.

\bibitem[{\citenamefont{Bennett and Brassard}(1984)}]{bennett1984quantum}
\bibinfo{author}{\bibfnamefont{C.~H.} \bibnamefont{Bennett}} \bibnamefont{and}
  \bibinfo{author}{\bibfnamefont{G.}~\bibnamefont{Brassard}}, in
  \emph{\bibinfo{booktitle}{Proceedings of the IEEE International Conference on
  Computers, Systems and Signal Processing}} (\bibinfo{publisher}{IEEE Press},
  \bibinfo{address}{New York}, \bibinfo{year}{1984}), pp.
  \bibinfo{pages}{175--179},
  \urlprefix\url{https://doi.org/10.1016/j.tcs.2014.05.025}.

\bibitem[{\citenamefont{Ekert}(1991)}]{ekert1991Quantum}
\bibinfo{author}{\bibfnamefont{A.~K.} \bibnamefont{Ekert}},
  \bibinfo{journal}{Phys. Rev. Lett.} \textbf{\bibinfo{volume}{67}},
  \bibinfo{pages}{661} (\bibinfo{year}{1991}),
  \urlprefix\url{http://link.aps.org/doi/10.1103/PhysRevLett.67.661}.

\bibitem[{\citenamefont{Giovannetti et~al.}(2004)\citenamefont{Giovannetti,
  Lloyd, and Maccone}}]{giovannetti2004quantum}
\bibinfo{author}{\bibfnamefont{V.}~\bibnamefont{Giovannetti}},
  \bibinfo{author}{\bibfnamefont{S.}~\bibnamefont{Lloyd}}, \bibnamefont{and}
  \bibinfo{author}{\bibfnamefont{L.}~\bibnamefont{Maccone}},
  \bibinfo{journal}{Science} \textbf{\bibinfo{volume}{306}},
  \bibinfo{pages}{1330} (\bibinfo{year}{2004}),
  \eprint{https://www.science.org/doi/pdf/10.1126/science.1104149},
  \urlprefix\url{https://www.science.org/doi/abs/10.1126/science.1104149}.

\bibitem[{\citenamefont{Giovannetti et~al.}(2011)\citenamefont{Giovannetti,
  Lloyd, and Maccone}}]{giovannetti2011advances}
\bibinfo{author}{\bibfnamefont{V.}~\bibnamefont{Giovannetti}},
  \bibinfo{author}{\bibfnamefont{S.}~\bibnamefont{Lloyd}}, \bibnamefont{and}
  \bibinfo{author}{\bibfnamefont{L.}~\bibnamefont{Maccone}},
  \bibinfo{journal}{Nature photonics} \textbf{\bibinfo{volume}{5}},
  \bibinfo{pages}{222} (\bibinfo{year}{2011}).

\bibitem[{\citenamefont{Chen et~al.}(2021)\citenamefont{Chen, Zhang, Chen, Cai,
  Liao, Zhang, Chen, Yin, Ren, Chen et~al.}}]{Chen2021integrated}
\bibinfo{author}{\bibfnamefont{Y.-A.} \bibnamefont{Chen}},
  \bibinfo{author}{\bibfnamefont{Q.}~\bibnamefont{Zhang}},
  \bibinfo{author}{\bibfnamefont{T.-Y.} \bibnamefont{Chen}},
  \bibinfo{author}{\bibfnamefont{W.-Q.} \bibnamefont{Cai}},
  \bibinfo{author}{\bibfnamefont{S.-K.} \bibnamefont{Liao}},
  \bibinfo{author}{\bibfnamefont{J.}~\bibnamefont{Zhang}},
  \bibinfo{author}{\bibfnamefont{K.}~\bibnamefont{Chen}},
  \bibinfo{author}{\bibfnamefont{J.}~\bibnamefont{Yin}},
  \bibinfo{author}{\bibfnamefont{J.-G.} \bibnamefont{Ren}},
  \bibinfo{author}{\bibfnamefont{Z.}~\bibnamefont{Chen}}, \bibnamefont{et~al.},
  \bibinfo{journal}{Nature} \textbf{\bibinfo{volume}{589}},
  \bibinfo{pages}{214} (\bibinfo{year}{2021}), ISSN \bibinfo{issn}{1476-4687},
  \urlprefix\url{https://doi.org/10.1038/s41586-020-03093-8}.

\bibitem[{\citenamefont{Peev et~al.}(2009)\citenamefont{Peev, Pacher,
  All{\'{e}}aume, Barreiro, Bouda, Boxleitner, Debuisschert, Diamanti, Dianati,
  Dynes et~al.}}]{peev2009secoqc}
\bibinfo{author}{\bibfnamefont{M.}~\bibnamefont{Peev}},
  \bibinfo{author}{\bibfnamefont{C.}~\bibnamefont{Pacher}},
  \bibinfo{author}{\bibfnamefont{R.}~\bibnamefont{All{\'{e}}aume}},
  \bibinfo{author}{\bibfnamefont{C.}~\bibnamefont{Barreiro}},
  \bibinfo{author}{\bibfnamefont{J.}~\bibnamefont{Bouda}},
  \bibinfo{author}{\bibfnamefont{W.}~\bibnamefont{Boxleitner}},
  \bibinfo{author}{\bibfnamefont{T.}~\bibnamefont{Debuisschert}},
  \bibinfo{author}{\bibfnamefont{E.}~\bibnamefont{Diamanti}},
  \bibinfo{author}{\bibfnamefont{M.}~\bibnamefont{Dianati}},
  \bibinfo{author}{\bibfnamefont{J.~F.} \bibnamefont{Dynes}},
  \bibnamefont{et~al.}, \bibinfo{journal}{New Journal of Physics}
  \textbf{\bibinfo{volume}{11}}, \bibinfo{pages}{075001}
  (\bibinfo{year}{2009}),
  \urlprefix\url{https://doi.org/10.1088/1367-2630/11/7/075001}.

\bibitem[{\citenamefont{Sasaki et~al.}(2011)\citenamefont{Sasaki, Fujiwara,
  Ishizuka, Klaus, Wakui, Takeoka, Miki, Yamashita, Wang, Tanaka
  et~al.}}]{sasaki2011field}
\bibinfo{author}{\bibfnamefont{M.}~\bibnamefont{Sasaki}},
  \bibinfo{author}{\bibfnamefont{M.}~\bibnamefont{Fujiwara}},
  \bibinfo{author}{\bibfnamefont{H.}~\bibnamefont{Ishizuka}},
  \bibinfo{author}{\bibfnamefont{W.}~\bibnamefont{Klaus}},
  \bibinfo{author}{\bibfnamefont{K.}~\bibnamefont{Wakui}},
  \bibinfo{author}{\bibfnamefont{M.}~\bibnamefont{Takeoka}},
  \bibinfo{author}{\bibfnamefont{S.}~\bibnamefont{Miki}},
  \bibinfo{author}{\bibfnamefont{T.}~\bibnamefont{Yamashita}},
  \bibinfo{author}{\bibfnamefont{Z.}~\bibnamefont{Wang}},
  \bibinfo{author}{\bibfnamefont{A.}~\bibnamefont{Tanaka}},
  \bibnamefont{et~al.}, \bibinfo{journal}{Opt. Express}
  \textbf{\bibinfo{volume}{19}}, \bibinfo{pages}{10387} (\bibinfo{year}{2011}),
  \urlprefix\url{http://opg.optica.org/oe/abstract.cfm?URI=oe-19-11-10387}.

\bibitem[{\citenamefont{Simon}(2017)}]{simon2017towards}
\bibinfo{author}{\bibfnamefont{C.}~\bibnamefont{Simon}},
  \bibinfo{journal}{Nature Photonics} \textbf{\bibinfo{volume}{11}},
  \bibinfo{pages}{678} (\bibinfo{year}{2017}).

\bibitem[{\citenamefont{Van~Meter and Devitt}(2016)}]{van2016path}
\bibinfo{author}{\bibfnamefont{R.}~\bibnamefont{Van~Meter}} \bibnamefont{and}
  \bibinfo{author}{\bibfnamefont{S.~J.} \bibnamefont{Devitt}},
  \bibinfo{journal}{Computer} \textbf{\bibinfo{volume}{49}},
  \bibinfo{pages}{31} (\bibinfo{year}{2016}).

\bibitem[{\citenamefont{Broadbent et~al.}(2009)\citenamefont{Broadbent,
  Fitzsimons, and Kashefi}}]{broadbent2009universal}
\bibinfo{author}{\bibfnamefont{A.}~\bibnamefont{Broadbent}},
  \bibinfo{author}{\bibfnamefont{J.}~\bibnamefont{Fitzsimons}},
  \bibnamefont{and} \bibinfo{author}{\bibfnamefont{E.}~\bibnamefont{Kashefi}},
  in \emph{\bibinfo{booktitle}{2009 50th annual IEEE symposium on foundations
  of computer science}} (\bibinfo{organization}{IEEE}, \bibinfo{year}{2009}),
  pp. \bibinfo{pages}{517--526}.

\bibitem[{\citenamefont{Briegel et~al.}(1998)\citenamefont{Briegel, D\"ur,
  Cirac, and Zoller}}]{Briegel1998Repeaters}
\bibinfo{author}{\bibfnamefont{H.-J.} \bibnamefont{Briegel}},
  \bibinfo{author}{\bibfnamefont{W.}~\bibnamefont{D\"ur}},
  \bibinfo{author}{\bibfnamefont{J.~I.} \bibnamefont{Cirac}}, \bibnamefont{and}
  \bibinfo{author}{\bibfnamefont{P.}~\bibnamefont{Zoller}},
  \bibinfo{journal}{Phys. Rev. Lett.} \textbf{\bibinfo{volume}{81}},
  \bibinfo{pages}{5932} (\bibinfo{year}{1998}),
  \urlprefix\url{https://link.aps.org/doi/10.1103/PhysRevLett.81.5932}.

\bibitem[{\citenamefont{Duan et~al.}(2001)\citenamefont{Duan, Lukin, Cirac, and
  Zoller}}]{duan2001long}
\bibinfo{author}{\bibfnamefont{L.-M.} \bibnamefont{Duan}},
  \bibinfo{author}{\bibfnamefont{M.}~\bibnamefont{Lukin}},
  \bibinfo{author}{\bibfnamefont{J.~I.} \bibnamefont{Cirac}}, \bibnamefont{and}
  \bibinfo{author}{\bibfnamefont{P.}~\bibnamefont{Zoller}},
  \bibinfo{journal}{Nature} \textbf{\bibinfo{volume}{414}},
  \bibinfo{pages}{413} (\bibinfo{year}{2001}).

\bibitem[{\citenamefont{Jacobs et~al.}(2002)\citenamefont{Jacobs, Pittman, and
  Franson}}]{jacobs2002quantum}
\bibinfo{author}{\bibfnamefont{B.~C.} \bibnamefont{Jacobs}},
  \bibinfo{author}{\bibfnamefont{T.~B.} \bibnamefont{Pittman}},
  \bibnamefont{and} \bibinfo{author}{\bibfnamefont{J.~D.}
  \bibnamefont{Franson}}, \bibinfo{journal}{Phys. Rev. A}
  \textbf{\bibinfo{volume}{66}}, \bibinfo{pages}{052307}
  (\bibinfo{year}{2002}),
  \urlprefix\url{https://link.aps.org/doi/10.1103/PhysRevA.66.052307}.

\bibitem[{\citenamefont{Waks et~al.}(2002)\citenamefont{Waks, Zeevi, and
  Yamamoto}}]{waks2002security}
\bibinfo{author}{\bibfnamefont{E.}~\bibnamefont{Waks}},
  \bibinfo{author}{\bibfnamefont{A.}~\bibnamefont{Zeevi}}, \bibnamefont{and}
  \bibinfo{author}{\bibfnamefont{Y.}~\bibnamefont{Yamamoto}},
  \bibinfo{journal}{Phys. Rev. A} \textbf{\bibinfo{volume}{65}},
  \bibinfo{pages}{052310} (\bibinfo{year}{2002}),
  \urlprefix\url{https://link.aps.org/doi/10.1103/PhysRevA.65.052310}.

\bibitem[{\citenamefont{de~Riedmatten et~al.}(2004)\citenamefont{de~Riedmatten,
  Marcikic, Tittel, Zbinden, Collins, and Gisin}}]{de2004long}
\bibinfo{author}{\bibfnamefont{H.}~\bibnamefont{de~Riedmatten}},
  \bibinfo{author}{\bibfnamefont{I.}~\bibnamefont{Marcikic}},
  \bibinfo{author}{\bibfnamefont{W.}~\bibnamefont{Tittel}},
  \bibinfo{author}{\bibfnamefont{H.}~\bibnamefont{Zbinden}},
  \bibinfo{author}{\bibfnamefont{D.}~\bibnamefont{Collins}}, \bibnamefont{and}
  \bibinfo{author}{\bibfnamefont{N.}~\bibnamefont{Gisin}},
  \bibinfo{journal}{Phys. Rev. Lett.} \textbf{\bibinfo{volume}{92}},
  \bibinfo{pages}{047904} (\bibinfo{year}{2004}),
  \urlprefix\url{https://link.aps.org/doi/10.1103/PhysRevLett.92.047904}.

\bibitem[{\citenamefont{Collins et~al.}(2005)\citenamefont{Collins, Gisin, and
  De~Riedmatten*}}]{collins2005quantum}
\bibinfo{author}{\bibfnamefont{D.}~\bibnamefont{Collins}},
  \bibinfo{author}{\bibfnamefont{N.}~\bibnamefont{Gisin}}, \bibnamefont{and}
  \bibinfo{author}{\bibfnamefont{H.}~\bibnamefont{De~Riedmatten*}},
  \bibinfo{journal}{Journal of Modern Optics} \textbf{\bibinfo{volume}{52}},
  \bibinfo{pages}{735} (\bibinfo{year}{2005}).

\bibitem[{\citenamefont{Jing et~al.}(2019)\citenamefont{Jing, Wang, Yu, Sun,
  Jiang, Yang, Jiang, Luo, Zhang, Jiang et~al.}}]{jing2019entanglement}
\bibinfo{author}{\bibfnamefont{B.}~\bibnamefont{Jing}},
  \bibinfo{author}{\bibfnamefont{X.-J.} \bibnamefont{Wang}},
  \bibinfo{author}{\bibfnamefont{Y.}~\bibnamefont{Yu}},
  \bibinfo{author}{\bibfnamefont{P.-F.} \bibnamefont{Sun}},
  \bibinfo{author}{\bibfnamefont{Y.}~\bibnamefont{Jiang}},
  \bibinfo{author}{\bibfnamefont{S.-J.} \bibnamefont{Yang}},
  \bibinfo{author}{\bibfnamefont{W.-H.} \bibnamefont{Jiang}},
  \bibinfo{author}{\bibfnamefont{X.-Y.} \bibnamefont{Luo}},
  \bibinfo{author}{\bibfnamefont{J.}~\bibnamefont{Zhang}},
  \bibinfo{author}{\bibfnamefont{X.}~\bibnamefont{Jiang}},
  \bibnamefont{et~al.}, \bibinfo{journal}{Nature Photonics}
  \textbf{\bibinfo{volume}{13}}, \bibinfo{pages}{210} (\bibinfo{year}{2019}).

\bibitem[{\citenamefont{Yu et~al.}(2020)\citenamefont{Yu, Ma, Luo, Jing, Sun,
  Fang, Yang, Liu, Zheng, Xie et~al.}}]{Yu2020Entanglement}
\bibinfo{author}{\bibfnamefont{Y.}~\bibnamefont{Yu}},
  \bibinfo{author}{\bibfnamefont{F.}~\bibnamefont{Ma}},
  \bibinfo{author}{\bibfnamefont{X.-Y.} \bibnamefont{Luo}},
  \bibinfo{author}{\bibfnamefont{B.}~\bibnamefont{Jing}},
  \bibinfo{author}{\bibfnamefont{P.-F.} \bibnamefont{Sun}},
  \bibinfo{author}{\bibfnamefont{R.-Z.} \bibnamefont{Fang}},
  \bibinfo{author}{\bibfnamefont{C.-W.} \bibnamefont{Yang}},
  \bibinfo{author}{\bibfnamefont{H.}~\bibnamefont{Liu}},
  \bibinfo{author}{\bibfnamefont{M.-Y.} \bibnamefont{Zheng}},
  \bibinfo{author}{\bibfnamefont{X.-P.} \bibnamefont{Xie}},
  \bibnamefont{et~al.}, \bibinfo{journal}{Nature}
  \textbf{\bibinfo{volume}{578}}, \bibinfo{pages}{240} (\bibinfo{year}{2020}),
  ISSN \bibinfo{issn}{1476-4687},
  \urlprefix\url{https://doi.org/10.1038/s41586-020-1976-7}.

\bibitem[{\citenamefont{Pompili et~al.}(2021)\citenamefont{Pompili, Hermans,
  Baier, Beukers, Humphreys, Schouten, Vermeulen, Tiggelman, dos
  Santos~Martins, Dirkse et~al.}}]{pompili2021realization}
\bibinfo{author}{\bibfnamefont{M.}~\bibnamefont{Pompili}},
  \bibinfo{author}{\bibfnamefont{S.~L.} \bibnamefont{Hermans}},
  \bibinfo{author}{\bibfnamefont{S.}~\bibnamefont{Baier}},
  \bibinfo{author}{\bibfnamefont{H.~K.} \bibnamefont{Beukers}},
  \bibinfo{author}{\bibfnamefont{P.~C.} \bibnamefont{Humphreys}},
  \bibinfo{author}{\bibfnamefont{R.~N.} \bibnamefont{Schouten}},
  \bibinfo{author}{\bibfnamefont{R.~F.} \bibnamefont{Vermeulen}},
  \bibinfo{author}{\bibfnamefont{M.~J.} \bibnamefont{Tiggelman}},
  \bibinfo{author}{\bibfnamefont{L.}~\bibnamefont{dos Santos~Martins}},
  \bibinfo{author}{\bibfnamefont{B.}~\bibnamefont{Dirkse}},
  \bibnamefont{et~al.}, \bibinfo{journal}{Science}
  \textbf{\bibinfo{volume}{372}}, \bibinfo{pages}{259} (\bibinfo{year}{2021}).

\bibitem[{\citenamefont{Hermans et~al.}(2022)\citenamefont{Hermans, Pompili,
  Beukers, Baier, Borregaard, and Hanson}}]{hermans2022qubit}
\bibinfo{author}{\bibfnamefont{S.}~\bibnamefont{Hermans}},
  \bibinfo{author}{\bibfnamefont{M.}~\bibnamefont{Pompili}},
  \bibinfo{author}{\bibfnamefont{H.}~\bibnamefont{Beukers}},
  \bibinfo{author}{\bibfnamefont{S.}~\bibnamefont{Baier}},
  \bibinfo{author}{\bibfnamefont{J.}~\bibnamefont{Borregaard}},
  \bibnamefont{and} \bibinfo{author}{\bibfnamefont{R.}~\bibnamefont{Hanson}},
  \bibinfo{journal}{Nature} \textbf{\bibinfo{volume}{605}},
  \bibinfo{pages}{663} (\bibinfo{year}{2022}).

\bibitem[{\citenamefont{Liu et~al.}(2024)\citenamefont{Liu, Luo, Yu, Wang,
  Wang, Hu, Li, Zheng, Yao, Yan et~al.}}]{liu2024creation}
\bibinfo{author}{\bibfnamefont{J.-L.} \bibnamefont{Liu}},
  \bibinfo{author}{\bibfnamefont{X.-Y.} \bibnamefont{Luo}},
  \bibinfo{author}{\bibfnamefont{Y.}~\bibnamefont{Yu}},
  \bibinfo{author}{\bibfnamefont{C.-Y.} \bibnamefont{Wang}},
  \bibinfo{author}{\bibfnamefont{B.}~\bibnamefont{Wang}},
  \bibinfo{author}{\bibfnamefont{Y.}~\bibnamefont{Hu}},
  \bibinfo{author}{\bibfnamefont{J.}~\bibnamefont{Li}},
  \bibinfo{author}{\bibfnamefont{M.-Y.} \bibnamefont{Zheng}},
  \bibinfo{author}{\bibfnamefont{B.}~\bibnamefont{Yao}},
  \bibinfo{author}{\bibfnamefont{Z.}~\bibnamefont{Yan}}, \bibnamefont{et~al.},
  \bibinfo{journal}{Nature} \textbf{\bibinfo{volume}{629}},
  \bibinfo{pages}{579} (\bibinfo{year}{2024}).

\bibitem[{\citenamefont{Lucamarini et~al.}(2018)\citenamefont{Lucamarini, Yuan,
  Dynes, and Shields}}]{lucamarini2018overcoming}
\bibinfo{author}{\bibfnamefont{M.}~\bibnamefont{Lucamarini}},
  \bibinfo{author}{\bibfnamefont{Z.}~\bibnamefont{Yuan}},
  \bibinfo{author}{\bibfnamefont{J.}~\bibnamefont{Dynes}}, \bibnamefont{and}
  \bibinfo{author}{\bibfnamefont{A.}~\bibnamefont{Shields}},
  \bibinfo{journal}{Nature} \textbf{\bibinfo{volume}{557}},
  \bibinfo{pages}{400} (\bibinfo{year}{2018}),
  \urlprefix\url{https://www.nature.com/articles/s41586-018-0066-6}.

\bibitem[{\citenamefont{Ma et~al.}(2018)\citenamefont{Ma, Zeng, and
  Zhou}}]{Ma2018phase}
\bibinfo{author}{\bibfnamefont{X.}~\bibnamefont{Ma}},
  \bibinfo{author}{\bibfnamefont{P.}~\bibnamefont{Zeng}}, \bibnamefont{and}
  \bibinfo{author}{\bibfnamefont{H.}~\bibnamefont{Zhou}},
  \bibinfo{journal}{Phys. Rev. X} \textbf{\bibinfo{volume}{8}},
  \bibinfo{pages}{031043} (\bibinfo{year}{2018}),
  \urlprefix\url{https://link.aps.org/doi/10.1103/PhysRevX.8.031043}.

\bibitem[{\citenamefont{Hwang}(2003)}]{hwang2003decoy}
\bibinfo{author}{\bibfnamefont{W.-Y.} \bibnamefont{Hwang}},
  \bibinfo{journal}{Phys. Rev. Lett.} \textbf{\bibinfo{volume}{91}},
  \bibinfo{pages}{057901} (\bibinfo{year}{2003}),
  \urlprefix\url{https://link.aps.org/doi/10.1103/PhysRevLett.91.057901}.

\bibitem[{\citenamefont{Lo et~al.}(2005)\citenamefont{Lo, Ma, and
  Chen}}]{Lo2005Decoy}
\bibinfo{author}{\bibfnamefont{H.-K.} \bibnamefont{Lo}},
  \bibinfo{author}{\bibfnamefont{X.}~\bibnamefont{Ma}}, \bibnamefont{and}
  \bibinfo{author}{\bibfnamefont{K.}~\bibnamefont{Chen}},
  \bibinfo{journal}{Phys. Rev. Lett.} \textbf{\bibinfo{volume}{94}},
  \bibinfo{pages}{230504} (\bibinfo{year}{2005}),
  \urlprefix\url{http://link.aps.org/doi/10.1103/PhysRevLett.94.230504}.

\bibitem[{\citenamefont{Wang}(2005)}]{wang2005decoy}
\bibinfo{author}{\bibfnamefont{X.-B.} \bibnamefont{Wang}},
  \bibinfo{journal}{Phys. Rev. Lett.} \textbf{\bibinfo{volume}{94}},
  \bibinfo{pages}{230503} (\bibinfo{year}{2005}),
  \urlprefix\url{https://link.aps.org/doi/10.1103/PhysRevLett.94.230503}.

\bibitem[{\citenamefont{Lo and Chau}(1999)}]{lo1999Unconditional}
\bibinfo{author}{\bibfnamefont{H.-K.} \bibnamefont{Lo}} \bibnamefont{and}
  \bibinfo{author}{\bibfnamefont{H.~F.} \bibnamefont{Chau}},
  \bibinfo{journal}{Science} \textbf{\bibinfo{volume}{283}},
  \bibinfo{pages}{2050} (\bibinfo{year}{1999}),
  \eprint{https://www.science.org/doi/pdf/10.1126/science.283.5410.2050},
  \urlprefix\url{https://www.science.org/doi/abs/10.1126/science.283.5410.2050}.

\bibitem[{\citenamefont{Kurtsiefer et~al.}(2000)\citenamefont{Kurtsiefer,
  Mayer, Zarda, and Weinfurter}}]{kurtsiefer2000stable}
\bibinfo{author}{\bibfnamefont{C.}~\bibnamefont{Kurtsiefer}},
  \bibinfo{author}{\bibfnamefont{S.}~\bibnamefont{Mayer}},
  \bibinfo{author}{\bibfnamefont{P.}~\bibnamefont{Zarda}}, \bibnamefont{and}
  \bibinfo{author}{\bibfnamefont{H.}~\bibnamefont{Weinfurter}},
  \bibinfo{journal}{Phys. Rev. Lett.} \textbf{\bibinfo{volume}{85}},
  \bibinfo{pages}{290} (\bibinfo{year}{2000}),
  \urlprefix\url{https://link.aps.org/doi/10.1103/PhysRevLett.85.290}.

\bibitem[{\citenamefont{Aharonovich et~al.}(2016)\citenamefont{Aharonovich,
  Englund, and Toth}}]{aharonovich2016solid}
\bibinfo{author}{\bibfnamefont{I.}~\bibnamefont{Aharonovich}},
  \bibinfo{author}{\bibfnamefont{D.}~\bibnamefont{Englund}}, \bibnamefont{and}
  \bibinfo{author}{\bibfnamefont{M.}~\bibnamefont{Toth}},
  \bibinfo{journal}{Nature photonics} \textbf{\bibinfo{volume}{10}},
  \bibinfo{pages}{631} (\bibinfo{year}{2016}).

\bibitem[{\citenamefont{Keller et~al.}(2004)\citenamefont{Keller, Lange,
  Hayasaka, Lange, and Walther}}]{keller2004continuous}
\bibinfo{author}{\bibfnamefont{M.}~\bibnamefont{Keller}},
  \bibinfo{author}{\bibfnamefont{B.}~\bibnamefont{Lange}},
  \bibinfo{author}{\bibfnamefont{K.}~\bibnamefont{Hayasaka}},
  \bibinfo{author}{\bibfnamefont{W.}~\bibnamefont{Lange}}, \bibnamefont{and}
  \bibinfo{author}{\bibfnamefont{H.}~\bibnamefont{Walther}},
  \bibinfo{journal}{Nature} \textbf{\bibinfo{volume}{431}},
  \bibinfo{pages}{1075} (\bibinfo{year}{2004}).

\bibitem[{\citenamefont{Michler et~al.}(2000)\citenamefont{Michler, Kiraz,
  Becher, Schoenfeld, Petroff, Zhang, Hu, and Imamoglu}}]{michler2000quantum}
\bibinfo{author}{\bibfnamefont{P.}~\bibnamefont{Michler}},
  \bibinfo{author}{\bibfnamefont{A.}~\bibnamefont{Kiraz}},
  \bibinfo{author}{\bibfnamefont{C.}~\bibnamefont{Becher}},
  \bibinfo{author}{\bibfnamefont{W.~V.} \bibnamefont{Schoenfeld}},
  \bibinfo{author}{\bibfnamefont{P.~M.} \bibnamefont{Petroff}},
  \bibinfo{author}{\bibfnamefont{L.}~\bibnamefont{Zhang}},
  \bibinfo{author}{\bibfnamefont{E.}~\bibnamefont{Hu}}, \bibnamefont{and}
  \bibinfo{author}{\bibfnamefont{A.}~\bibnamefont{Imamoglu}},
  \bibinfo{journal}{Science} \textbf{\bibinfo{volume}{290}},
  \bibinfo{pages}{2282} (\bibinfo{year}{2000}),
  \eprint{https://www.science.org/doi/pdf/10.1126/science.290.5500.2282},
  \urlprefix\url{https://www.science.org/doi/abs/10.1126/science.290.5500.2282}.

\bibitem[{\citenamefont{Ding et~al.}(2016)\citenamefont{Ding, He, Duan,
  Gregersen, Chen, Unsleber, Maier, Schneider, Kamp, H\"ofling
  et~al.}}]{ding2016demand}
\bibinfo{author}{\bibfnamefont{X.}~\bibnamefont{Ding}},
  \bibinfo{author}{\bibfnamefont{Y.}~\bibnamefont{He}},
  \bibinfo{author}{\bibfnamefont{Z.-C.} \bibnamefont{Duan}},
  \bibinfo{author}{\bibfnamefont{N.}~\bibnamefont{Gregersen}},
  \bibinfo{author}{\bibfnamefont{M.-C.} \bibnamefont{Chen}},
  \bibinfo{author}{\bibfnamefont{S.}~\bibnamefont{Unsleber}},
  \bibinfo{author}{\bibfnamefont{S.}~\bibnamefont{Maier}},
  \bibinfo{author}{\bibfnamefont{C.}~\bibnamefont{Schneider}},
  \bibinfo{author}{\bibfnamefont{M.}~\bibnamefont{Kamp}},
  \bibinfo{author}{\bibfnamefont{S.}~\bibnamefont{H\"ofling}},
  \bibnamefont{et~al.}, \bibinfo{journal}{Phys. Rev. Lett.}
  \textbf{\bibinfo{volume}{116}}, \bibinfo{pages}{020401}
  (\bibinfo{year}{2016}),
  \urlprefix\url{https://link.aps.org/doi/10.1103/PhysRevLett.116.020401}.

\bibitem[{\citenamefont{Ding et~al.}(2025)\citenamefont{Ding, Guo, Xu, Liu,
  Zou, Zhao, Ge, Zhang, Liu, Wang et~al.}}]{ding2023highefficiency}
\bibinfo{author}{\bibfnamefont{X.}~\bibnamefont{Ding}},
  \bibinfo{author}{\bibfnamefont{Y.-P.} \bibnamefont{Guo}},
  \bibinfo{author}{\bibfnamefont{M.-C.} \bibnamefont{Xu}},
  \bibinfo{author}{\bibfnamefont{R.-Z.} \bibnamefont{Liu}},
  \bibinfo{author}{\bibfnamefont{G.-Y.} \bibnamefont{Zou}},
  \bibinfo{author}{\bibfnamefont{J.-Y.} \bibnamefont{Zhao}},
  \bibinfo{author}{\bibfnamefont{Z.-X.} \bibnamefont{Ge}},
  \bibinfo{author}{\bibfnamefont{Q.-H.} \bibnamefont{Zhang}},
  \bibinfo{author}{\bibfnamefont{H.-L.} \bibnamefont{Liu}},
  \bibinfo{author}{\bibfnamefont{L.-J.} \bibnamefont{Wang}},
  \bibnamefont{et~al.}, \bibinfo{journal}{Nature Photonics}
  \textbf{\bibinfo{volume}{19}}, \bibinfo{pages}{387} (\bibinfo{year}{2025}),
  ISSN \bibinfo{issn}{1749-4893},
  \urlprefix\url{https://doi.org/10.1038/s41566-025-01639-8}.

\bibitem[{\citenamefont{Chou et~al.}(2004)\citenamefont{Chou, Polyakov,
  Kuzmich, and Kimble}}]{chou2004single}
\bibinfo{author}{\bibfnamefont{C.~W.} \bibnamefont{Chou}},
  \bibinfo{author}{\bibfnamefont{S.~V.} \bibnamefont{Polyakov}},
  \bibinfo{author}{\bibfnamefont{A.}~\bibnamefont{Kuzmich}}, \bibnamefont{and}
  \bibinfo{author}{\bibfnamefont{H.~J.} \bibnamefont{Kimble}},
  \bibinfo{journal}{Phys. Rev. Lett.} \textbf{\bibinfo{volume}{92}},
  \bibinfo{pages}{213601} (\bibinfo{year}{2004}),
  \urlprefix\url{https://link.aps.org/doi/10.1103/PhysRevLett.92.213601}.

\bibitem[{\citenamefont{Ripka et~al.}(2018)\citenamefont{Ripka, K{\"u}bler,
  L{\"o}w, and Pfau}}]{ripka2018room}
\bibinfo{author}{\bibfnamefont{F.}~\bibnamefont{Ripka}},
  \bibinfo{author}{\bibfnamefont{H.}~\bibnamefont{K{\"u}bler}},
  \bibinfo{author}{\bibfnamefont{R.}~\bibnamefont{L{\"o}w}}, \bibnamefont{and}
  \bibinfo{author}{\bibfnamefont{T.}~\bibnamefont{Pfau}},
  \bibinfo{journal}{Science} \textbf{\bibinfo{volume}{362}},
  \bibinfo{pages}{446} (\bibinfo{year}{2018}).

\bibitem[{\citenamefont{Zeng et~al.}(2020)\citenamefont{Zeng, Wu, and
  Ma}}]{Zeng2019Symmetryprotected}
\bibinfo{author}{\bibfnamefont{P.}~\bibnamefont{Zeng}},
  \bibinfo{author}{\bibfnamefont{W.}~\bibnamefont{Wu}}, \bibnamefont{and}
  \bibinfo{author}{\bibfnamefont{X.}~\bibnamefont{Ma}}, \bibinfo{journal}{Phys.
  Rev. Appl.} \textbf{\bibinfo{volume}{13}}, \bibinfo{pages}{064013}
  (\bibinfo{year}{2020}),
  \urlprefix\url{https://link.aps.org/doi/10.1103/PhysRevApplied.13.064013}.

\bibitem[{\citenamefont{Huang et~al.}(2023)\citenamefont{Huang, Du, and
  Ma}}]{huang2023source}
\bibinfo{author}{\bibfnamefont{Y.}~\bibnamefont{Huang}},
  \bibinfo{author}{\bibfnamefont{Z.}~\bibnamefont{Du}}, \bibnamefont{and}
  \bibinfo{author}{\bibfnamefont{X.}~\bibnamefont{Ma}},
  \bibinfo{journal}{Advanced Quantum Technologies} p. \bibinfo{pages}{2300275}
  (\bibinfo{year}{2023}),
  \eprint{https://onlinelibrary.wiley.com/doi/pdf/10.1002/qute.202300275},
  \urlprefix\url{https://onlinelibrary.wiley.com/doi/abs/10.1002/qute.202300275}.

\bibitem[{\citenamefont{Glauber}(1963)}]{glauber1963quantum}
\bibinfo{author}{\bibfnamefont{R.~J.} \bibnamefont{Glauber}},
  \bibinfo{journal}{Physical Review} \textbf{\bibinfo{volume}{130}},
  \bibinfo{pages}{2529} (\bibinfo{year}{1963}).

\bibitem[{\citenamefont{Deng et~al.}(2019)\citenamefont{Deng, Wang, Ding, Duan,
  Qin, Chen, He, He, Li, Li et~al.}}]{Deng2019quantum}
\bibinfo{author}{\bibfnamefont{Y.-H.} \bibnamefont{Deng}},
  \bibinfo{author}{\bibfnamefont{H.}~\bibnamefont{Wang}},
  \bibinfo{author}{\bibfnamefont{X.}~\bibnamefont{Ding}},
  \bibinfo{author}{\bibfnamefont{Z.-C.} \bibnamefont{Duan}},
  \bibinfo{author}{\bibfnamefont{J.}~\bibnamefont{Qin}},
  \bibinfo{author}{\bibfnamefont{M.-C.} \bibnamefont{Chen}},
  \bibinfo{author}{\bibfnamefont{Y.}~\bibnamefont{He}},
  \bibinfo{author}{\bibfnamefont{Y.-M.} \bibnamefont{He}},
  \bibinfo{author}{\bibfnamefont{J.-P.} \bibnamefont{Li}},
  \bibinfo{author}{\bibfnamefont{Y.-H.} \bibnamefont{Li}},
  \bibnamefont{et~al.}, \bibinfo{journal}{Phys. Rev. Lett.}
  \textbf{\bibinfo{volume}{123}}, \bibinfo{pages}{080401}
  (\bibinfo{year}{2019}),
  \urlprefix\url{https://link.aps.org/doi/10.1103/PhysRevLett.123.080401}.

\bibitem[{\citenamefont{Minder et~al.}(2019)\citenamefont{Minder, Pittaluga,
  Roberts, Lucamarini, Dynes, Yuan, and Shields}}]{minder2019experimental}
\bibinfo{author}{\bibfnamefont{M.}~\bibnamefont{Minder}},
  \bibinfo{author}{\bibfnamefont{M.}~\bibnamefont{Pittaluga}},
  \bibinfo{author}{\bibfnamefont{G.~L.} \bibnamefont{Roberts}},
  \bibinfo{author}{\bibfnamefont{M.}~\bibnamefont{Lucamarini}},
  \bibinfo{author}{\bibfnamefont{J.~F.} \bibnamefont{Dynes}},
  \bibinfo{author}{\bibfnamefont{Z.~L.} \bibnamefont{Yuan}}, \bibnamefont{and}
  \bibinfo{author}{\bibfnamefont{A.~J.} \bibnamefont{Shields}},
  \bibinfo{journal}{Nature Photonics} \textbf{\bibinfo{volume}{13}},
  \bibinfo{pages}{334} (\bibinfo{year}{2019}), ISSN \bibinfo{issn}{1749-4893},
  \urlprefix\url{https://doi.org/10.1038/s41566-019-0377-7}.

\end{thebibliography}
\end{document}